\documentclass[lettersize,journal]{IEEEtran}
\usepackage{amsmath,amsfonts}
\usepackage{textcomp}
\usepackage{stfloats}
\usepackage{url}
\usepackage{verbatim}
\usepackage{graphicx}
\usepackage{cite}
\usepackage{amsmath,amsfonts}
\usepackage{algorithmic}
\usepackage{algorithm}
\usepackage{array}
\usepackage[caption=false,font=footnotesize,labelfont=rm,textfont=rm]{subfig}
\usepackage{bbding}
\usepackage{booktabs}
\usepackage{multirow}
\usepackage{makecell}
\usepackage{amsmath,amsthm,amssymb,amsfonts}
\usepackage{threeparttable}
\usepackage{xcolor} 

\begin{document}
\title{iADCPS: Time Series Anomaly Detection for Evolving Cyber-physical Systems via Incremental Meta-learning}

\author{Jiyu Tian,~Mingchu Li,~Liming Chen, ~\IEEEmembership{Senior Member,~IEEE},~Zumin Wang
\thanks{This work has been submitted to the IEEE for possible publication. Copyright may be transferred without notice, after which this version may no longer be accessible.}

\thanks{Jiyu Tian and Mingchu Li are with the School of Software Technology, Dalian University of Technology, Dalian, China. Mingchu Li is also affiliated with the School of Computer and Information Engineering, Jiangxi Normal University, Nanchang, Jiangxi, China. Email: tianjiyu@mail.dlut.edu.cn, mingchul@dlut.edu.cn}

\thanks{Liming Chen is with the School of Computer Science and Technology, Dalian University of Technology, Dalian, China. Email: limingchen0922@dlut.edu.cn}

\thanks{Zumin Wang is with the College of Information Engineering, Dalian University, Dalian, China. Email: wangzumin@dlu.edu.cn}


}
\markboth{arxiv }
{How to Use the IEEEtran \LaTeX \ Templates}

\maketitle

\begin{abstract}
Anomaly detection for cyber-physical systems (ADCPS) is crucial in identifying faults and potential attacks by analyzing the time series of sensor measurements and actuator states. However, current methods lack adaptation to data distribution shifts in both temporal and spatial dimensions as cyber-physical systems evolve. To tackle this issue, we propose an incremental meta-learning-based approach, namely iADCPS, which can continuously update the model through limited evolving normal samples to reconcile the distribution gap between evolving and historical time series. Specifically, We first introduce a temporal mixup strategy to align data for data-level generalization which is then combined with the one-class meta-learning approach for model-level generalization. Furthermore, we develop a non-parametric dynamic threshold to adaptively adjust the threshold based on the probability density of the abnormal scores without any anomaly supervision.
We empirically evaluate the effectiveness of the iADCPS using three publicly available datasets PUMP, SWaT, and WADI. The experimental results demonstrate that our method achieves 99.0\%, 93.1\%, and 78.7\% F1-Score, respectively, which outperforms the state-of-the-art (SOTA) ADCPS method, especially in the context of the evolving CPSs.

\end{abstract}


\section{Introduction}
\label{sec:1}
\begin{figure}[h]
  \centering
  \includegraphics[width=0.88\linewidth]{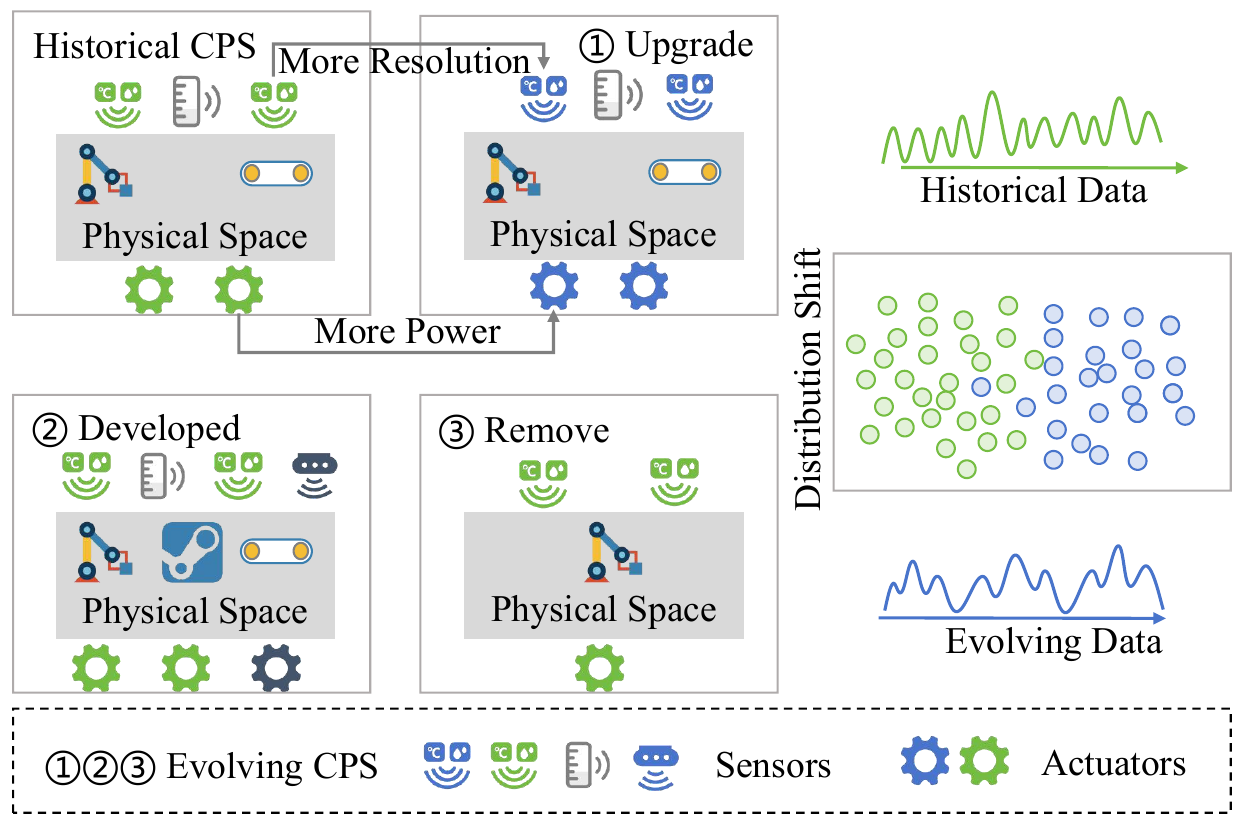}
  \caption{An illustrative example of CPS evolution.}
  \label{figure1}
\end{figure}
\IEEEPARstart{C}{yber}-physical systems (CPS) have become essential in critical infrastructures such as intelligent transportation, smart healthcare, and industrial control, which effectively enhance operational efficiency and automation within these sectors \cite{survey1, survey2,survey3,dctgan,amcnnlstm}. However, the complex interconnections within CPSs render them susceptible to anomalies, ranging from physical and digital component failures to cyber-attacks. In CPS, sensors and actuators play pivotal roles in tracking the system states over time, saving the collected measurements and states as time series. Consequently, considerable attention has been directed towards time series-based ADCPS. These approaches aim to pinpoint the precise instants of anomalies and failures by detecting deviations in time-series patterns, thereby aiding in prompt response strategies, including both the static-based methods \cite{lstm_vae,cgcd,nsibf,fusagnet} and the dynamic-based methods\cite{lstm_ndt, arcus, acudl, adt}.

However, the rapid and ongoing development of CPS presents significant challenges for existing detection methods. With sensor and actuator technologies undergoing constant upgrades and changing business needs, the time-series patterns of CPS systems can undergo substantial transformations. Figure \ref{figure1} illustrates three typical scenarios of CPS system evolution: i) the replacement of current devices due to system upgrades; ii) the increase of current devices resulting from system expansion; and iii) the removal of current devices owing to system reconfiguration.

Despite the increasing attention on CPS evolution, existing anomaly detection models fail to satisfy the performance requirements for the following reasons: ADCPS methods identify anomalies by detecting data points that deviate beyond thresholds from normal operational ranges. However, static approaches struggle to adapt to frequently changing time-series patterns as the models are trained on fixed distributions and the predefined thresholds prove challenging to adjust for unstable data; on the other hand, dynamic methods, while capable of adjusting thresholds with evolving data, still rely on hard-to-obtain anomaly labels. As such, existing ADCPS methods still face the following challenges:

\begin{itemize}
\item \textbf{Generalization:} Existing ADCPS methods constrain the effective generalization of models on continuously evolving time series. For example, NSIBF \cite{nsibf} captures CPS anomaly patterns using an end-to-end state-space model (SSM) \cite{ssmkf}, but the static SSM trained on historical data struggles to adapt to evolving data. ACUDL \cite{acudl} introduces a dynamic graph mechanism to achieve unsupervised adaptive updating of the model but ignores the distribution alignment and the reliance on large-scale normal samples for frequent incremental training. Failure to leverage limited evolving normal samples for incremental training impedes the detection model from responding effectively to CPS evolution.

\item \textbf{Adaptive threshold:} Existing ADCPS methods encounter challenges in flexibly adjusting thresholds alongside model updates. For instance, while ADT \cite{adt} employs a reinforcement learning mechanism to dynamically adjust the threshold, but necessitates frequent supervision inputs (less than 50 timestamps) to determine system states as normal or abnormal. LSTM-NDT \cite{lstm_ndt} calculates thresholds based on the differences between the abnormal series and the original series through an anomaly pruning strategy, that still relies on a limited number of anomalous labels. When the threshold fails to adapt suitably to evolving data, maintaining optimal detection performance for anomalies becomes a significant hurdle.
\end{itemize}

To overcome the challenges above, we propose the iADCPS, a novel approach that integrates incremental training and meta-learning methods to address distribution shifts resulting from CPS evolution. Specifically, we introduce a dual-adaptive incremental training framework (dual-adapter) tailored for CPS time series to address the challenge of generalization, which mixes historical and evolving data through the temporal Mixup algorithm for data-level adaptation and combines it with the one-class meta-learning approach for model-level adaptation. Furthermore, we present a non-parametric dynamic threshold based on low-density points (LDP-DT), which achieves adaptive threshold adjustments based on the probability density distribution of abnormal scores without any preset and anomaly labels. Unlike previous studies on CPS anomaly detection, our dual-adaptive training framework is based on both data and model levels to generalize evolving time series with minimal normal samples in each incremental task, while the dynamic threshold algorithm based on low-density points eliminates the need for anomaly labels while ensuring detection accuracy. We validate our approach through comprehensive qualitative and quantitative experiments conducted on simulated data and three real CPS datasets. The experimental results demonstrate that our method outperforms SOTA time series anomaly detection methods in both stable and evolving CPS scenarios. We summarise the main contributions of this paper as follows:

\begin{itemize}
\item We propose Dual-Adapter, an incremental meta-learning framework for anomaly detection in evolving CPS, where data alignment and meta-training enable the detection model to achieve effective generalization using only a few normal samples in the evolving data.
\item We design a non-parametric dynamic thresholding method based on low-density points to address the problem of adaptive threshold adjustment during incremental model updates.
\item We show that our proposed method consistently outperforms the SOTA in both stable and evolved CPS anomaly detection by extensive qualitative and quantitative experiments.
\end{itemize}

\section{Background and Motivation}
\label{sec:2}
\subsection{Time Series Anomaly Detection for Cyber-physical Systems}
\label{sec:2.1}

A typical CPS structure usually contains three types of devices: sensors, actuators, and programmable logic controllers (PLCs). The sensors convert physical parameters into electronic measurements; the PLC sends control commands to the actuators based on the measurements received from the sensors; and the actuators convert these control commands into physical state changes (e.g., opening or closing a valve). 

\begin{figure}[h]
  \centering
  \includegraphics[width=0.88\linewidth]{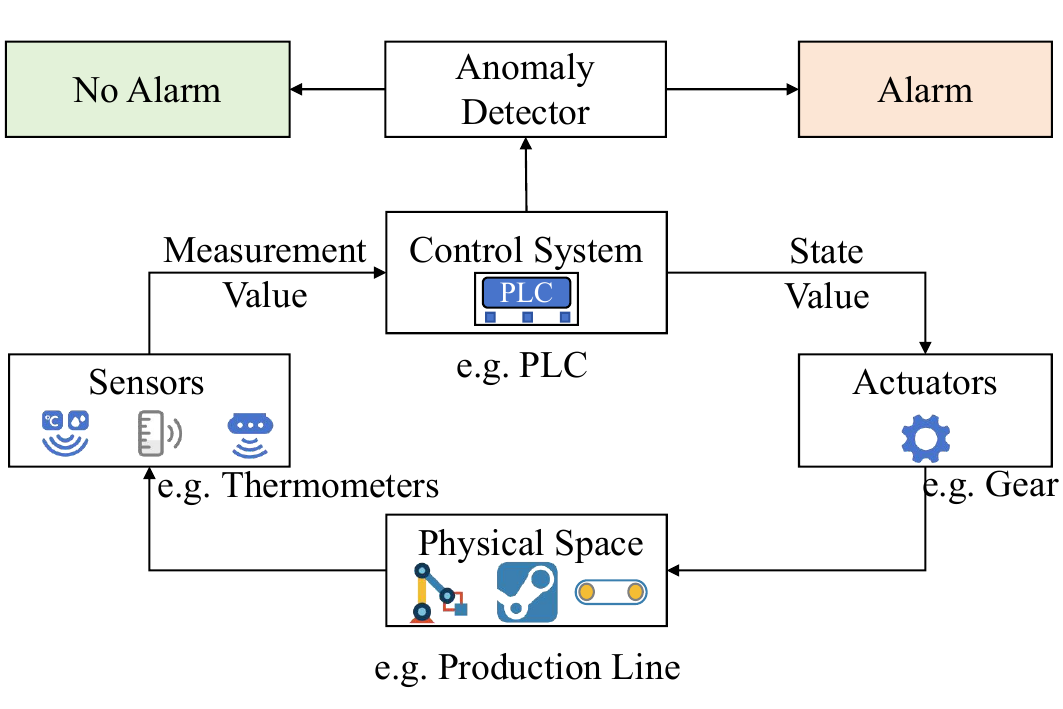}
  \caption{An illustrative example of ADCPS.}
  \label{figure2}
\end{figure}

To protect CPS security and stability, time series-based ADCPS methods rely on sensor measurements and actuator states generated by the above devices to capture anomalies. These methods typically fall into two categories: residual error-based and density-based methods. Residual error-based approaches use the prediction model \cite{lstm_ndt,lstm_vae} (e.g., recurrent neural network) or the reconstruction model \cite{adt, ccsae, aecid} (e.g., autoencoder) to calculate differences between predicted/reconstructed and observed values, flagging anomalies when the residuals surpass a threshold. However, such methods are susceptible to noise because defining precise thresholds for residual errors is challenging. In contrast, density-based detection methods combine SSM \cite{ssmkf} and Kalman filtering \cite{kalmanfilter} to robustly track and predict the system state in noisy environments, thus effectively resisting the effects of noise from sensors, actuators, and processes.

Among cutting-edge density-based detection methods, NSIBF \cite{nsibf} demonstrates advanced detection accuracy and potential noise immunity compared to residual error detection methods. Meanwhile, the end-to-end architecture of NSIBF reduces reliance on prior system dynamics knowledge, setting it apart from traditional density detection methods. Inspired by NSIBF, our study enhances this approach by replacing the conventional SSM with three sub-neural networks $\textit{f}_{net}$, $\textit{g}_{net}$, and $\textit{h}_{net}$ for end-to-end adaptive updating, which potentially preserves the robustness of the detection model while eliminating the need to rely on prior knowledge.

However, the static nature of NSIBF, reliant on fixed distributions and predefined thresholds, hinders its adaptability to evolving CPS environments and limits its overall generalizability. This rigidity poses a challenge as CPS systems evolve, highlighting the necessity for more flexible and adaptive anomaly detection strategies.

\subsection{Cyber-physical Systems Evolution}
\label{sec:2.2}
\begin{figure*}[!htb]

	\centering
	\scriptsize
	\subfloat[PUMP-Original]{
		\includegraphics[width=0.235\linewidth]{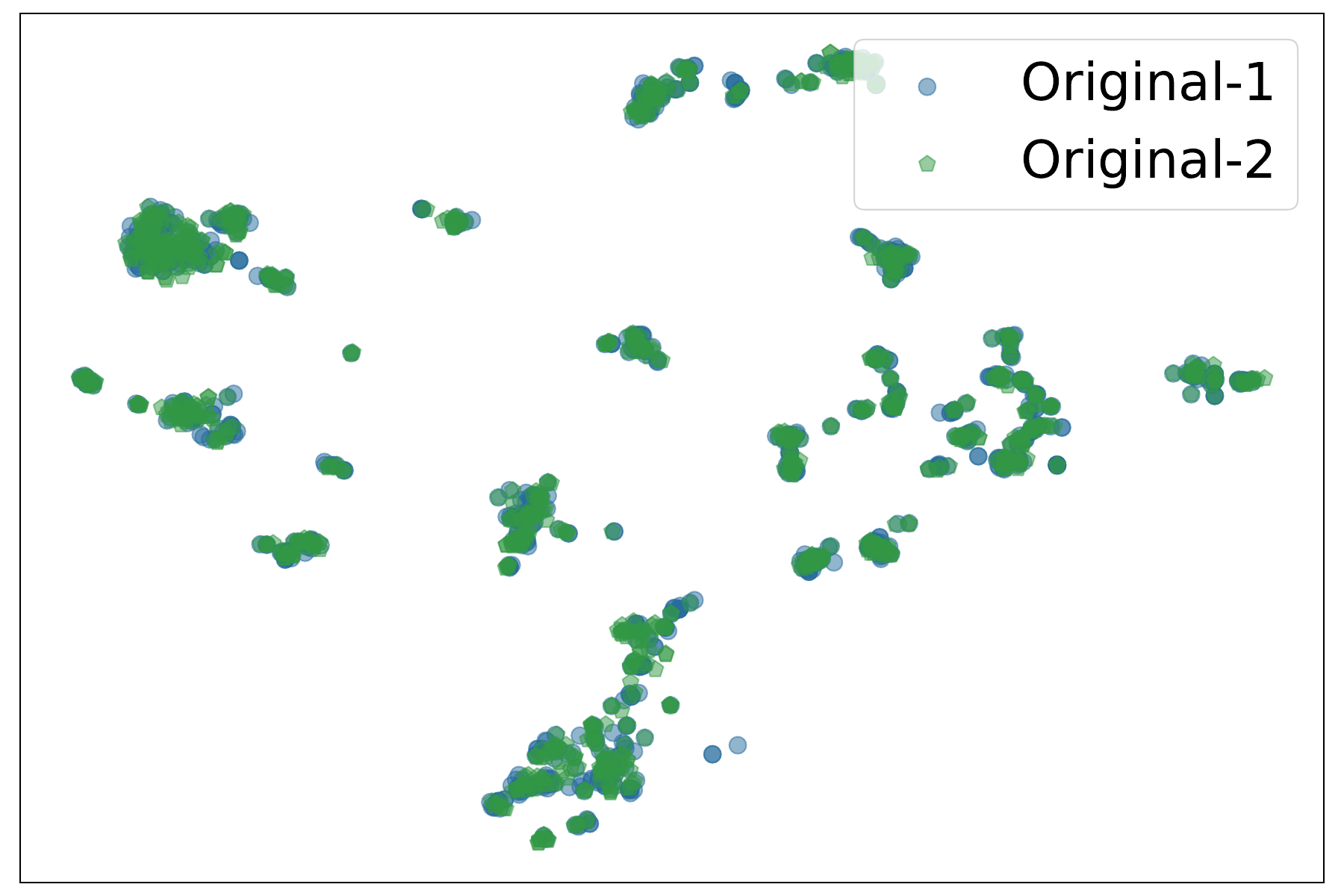}
	}
	\hfill
	\subfloat[PUMP-Remove]{
		\includegraphics[width=0.235\linewidth]{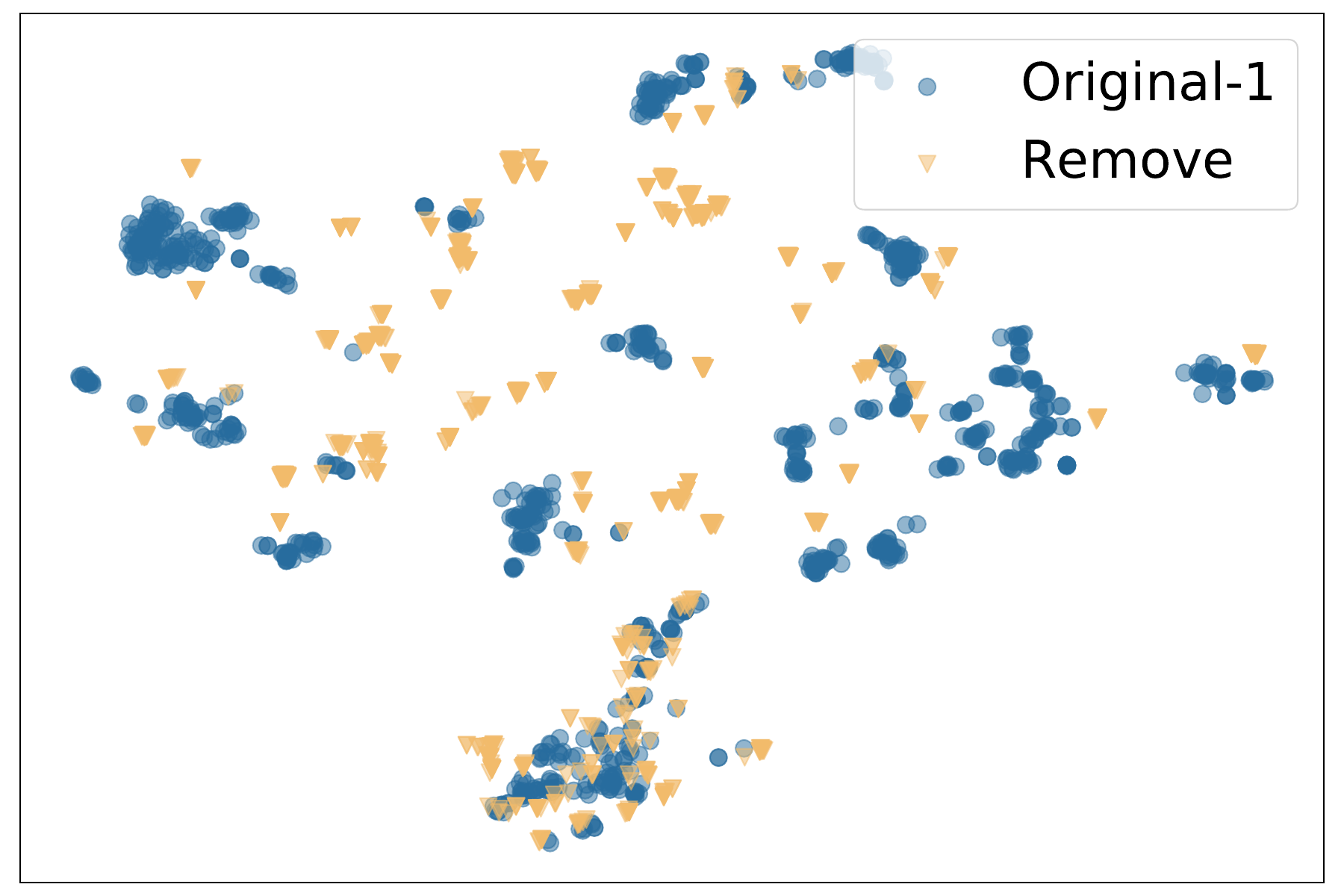}
	}
	\hfill
	\subfloat[PUMP-Upgrade]{
		\includegraphics[width=0.235\linewidth]{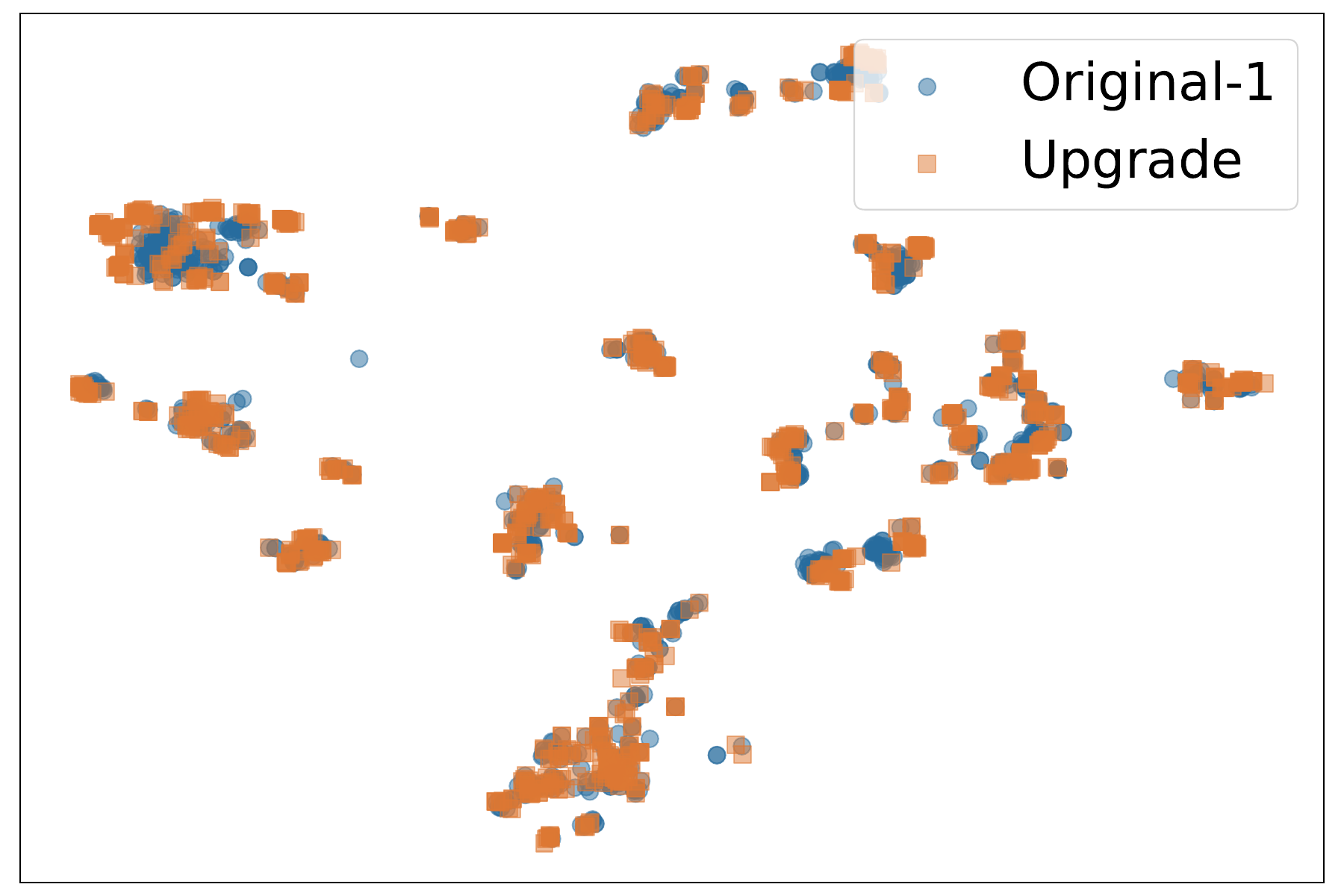}
	}
	\hfill
	\subfloat[PUMP-Mix]{
		\includegraphics[width=0.235\linewidth]{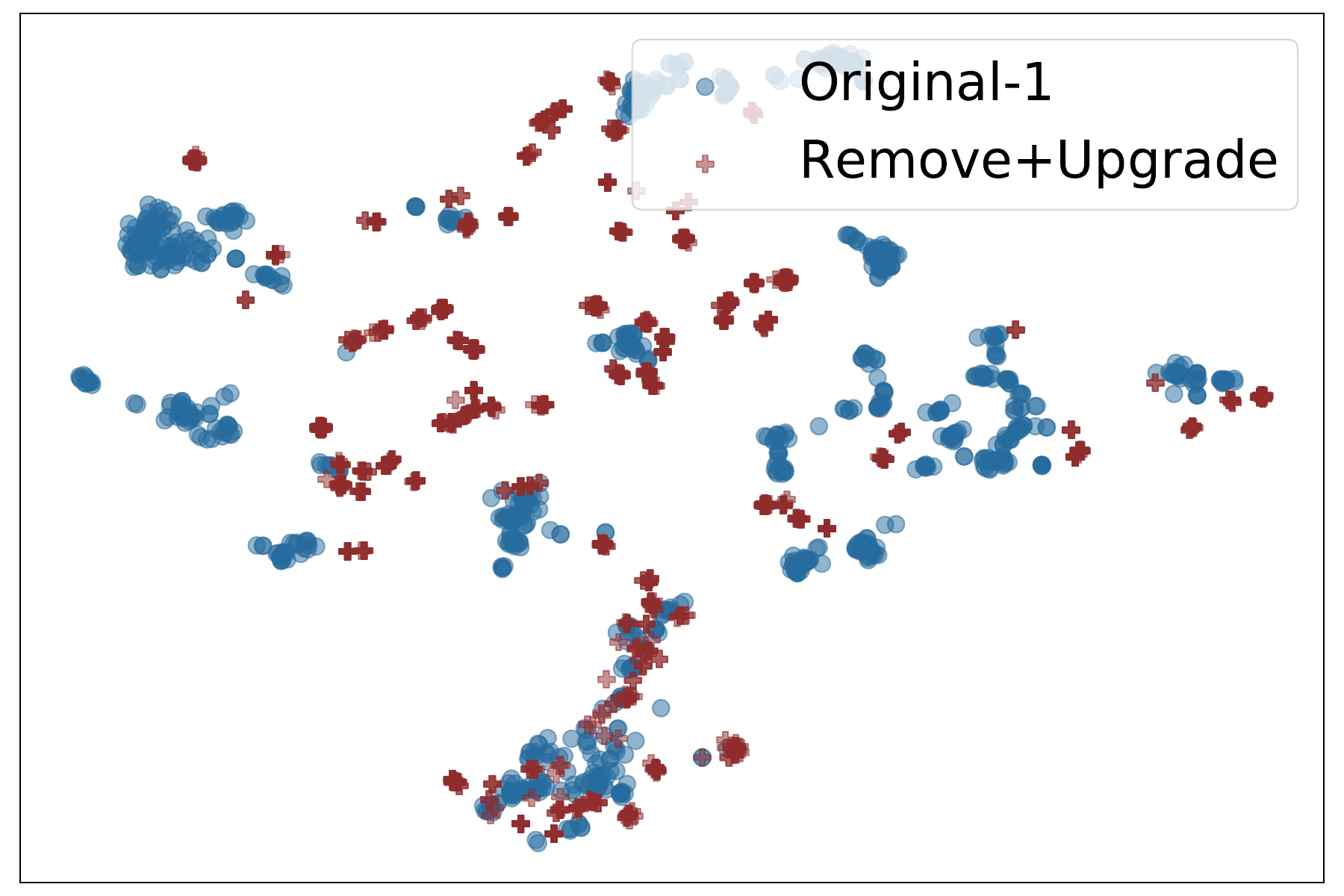}
	}
 
        \centering
        \scriptsize
 	\hfill
	\subfloat[SWaT-Original]{
		\includegraphics[width=0.235\linewidth]{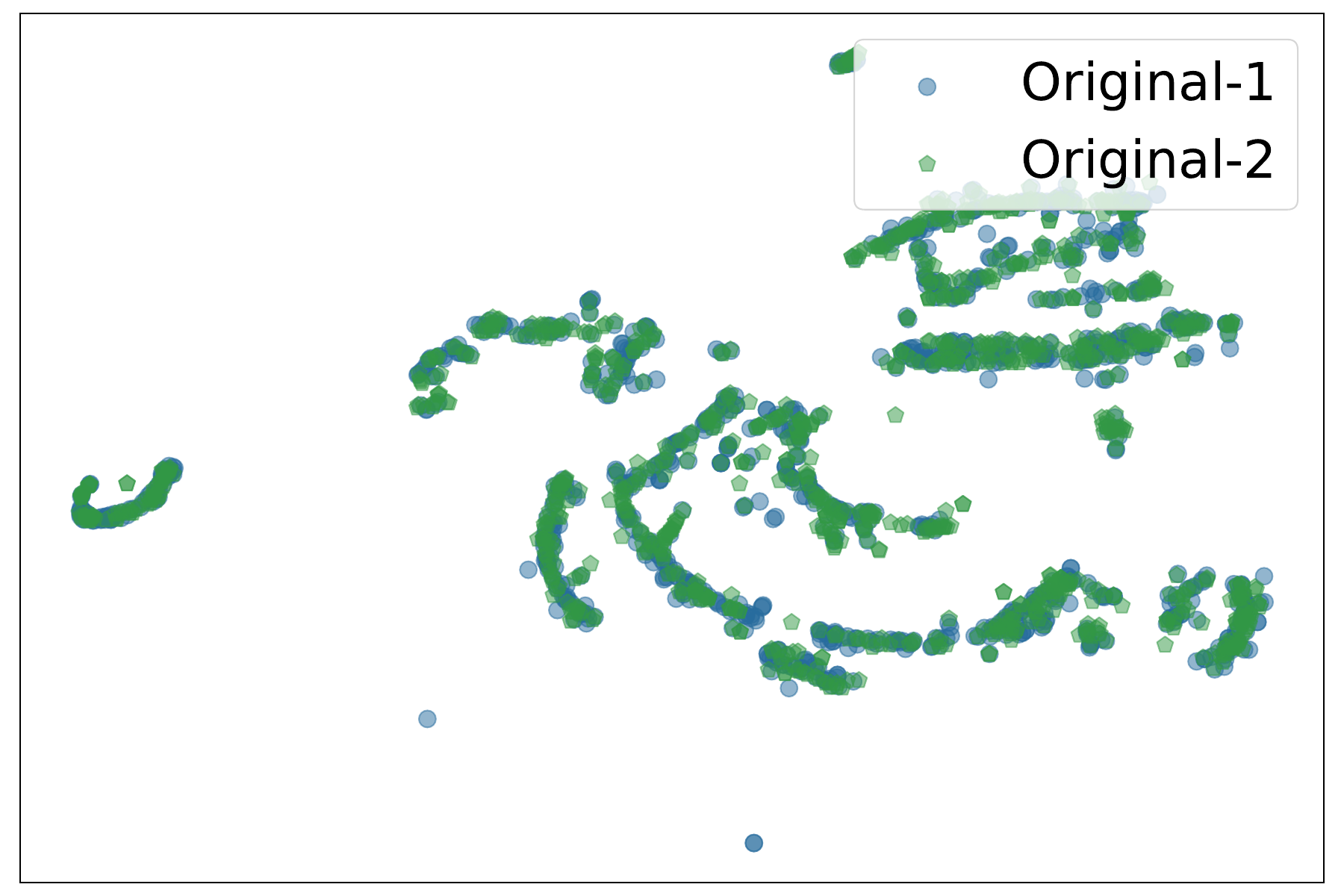}
	}
	\hfill
	\subfloat[SWaT-Remove]{
		\includegraphics[width=0.235\linewidth]{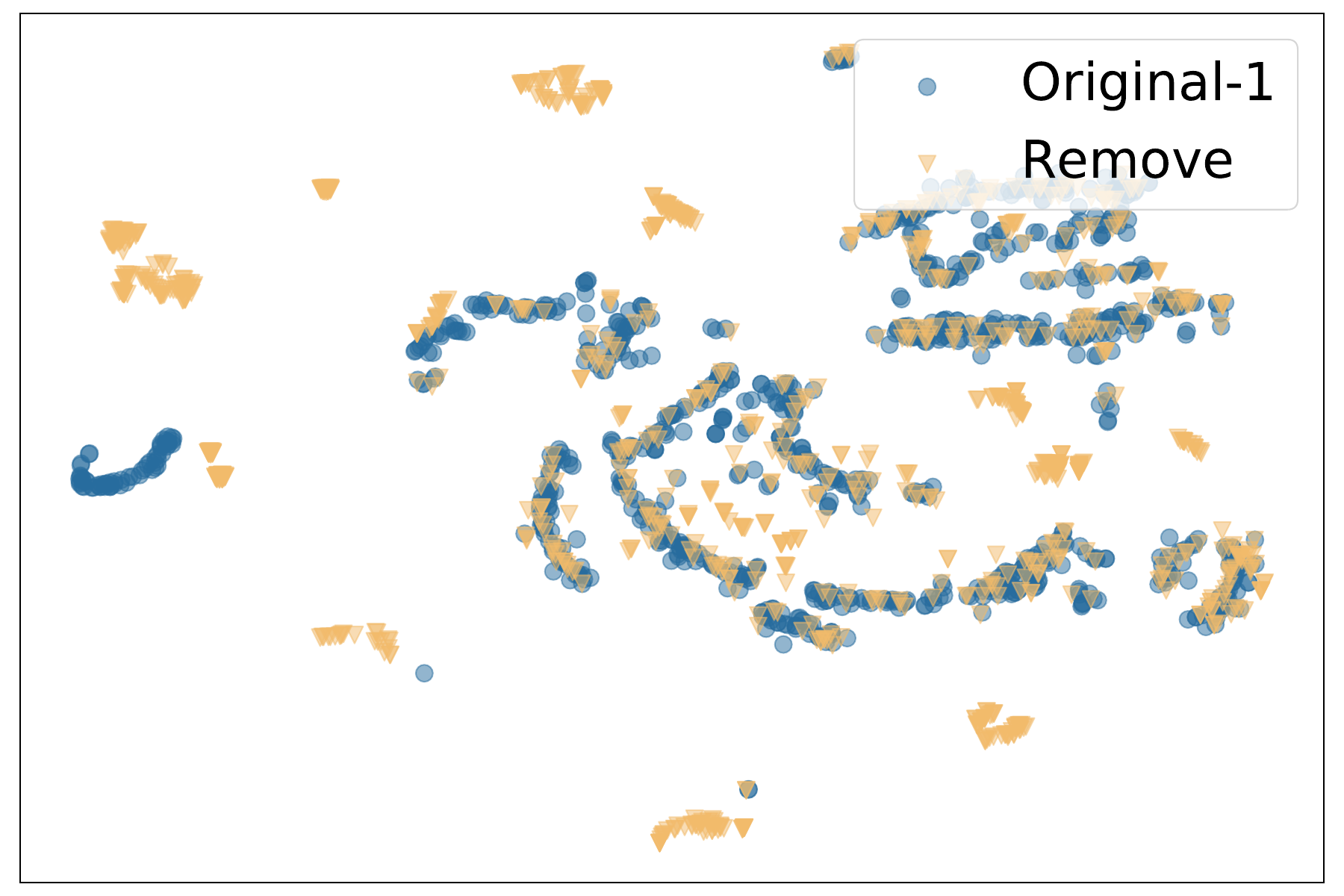}
	}
	\scriptsize
	\subfloat[SWaT-Upgrade]{
		\includegraphics[width=0.235\linewidth]{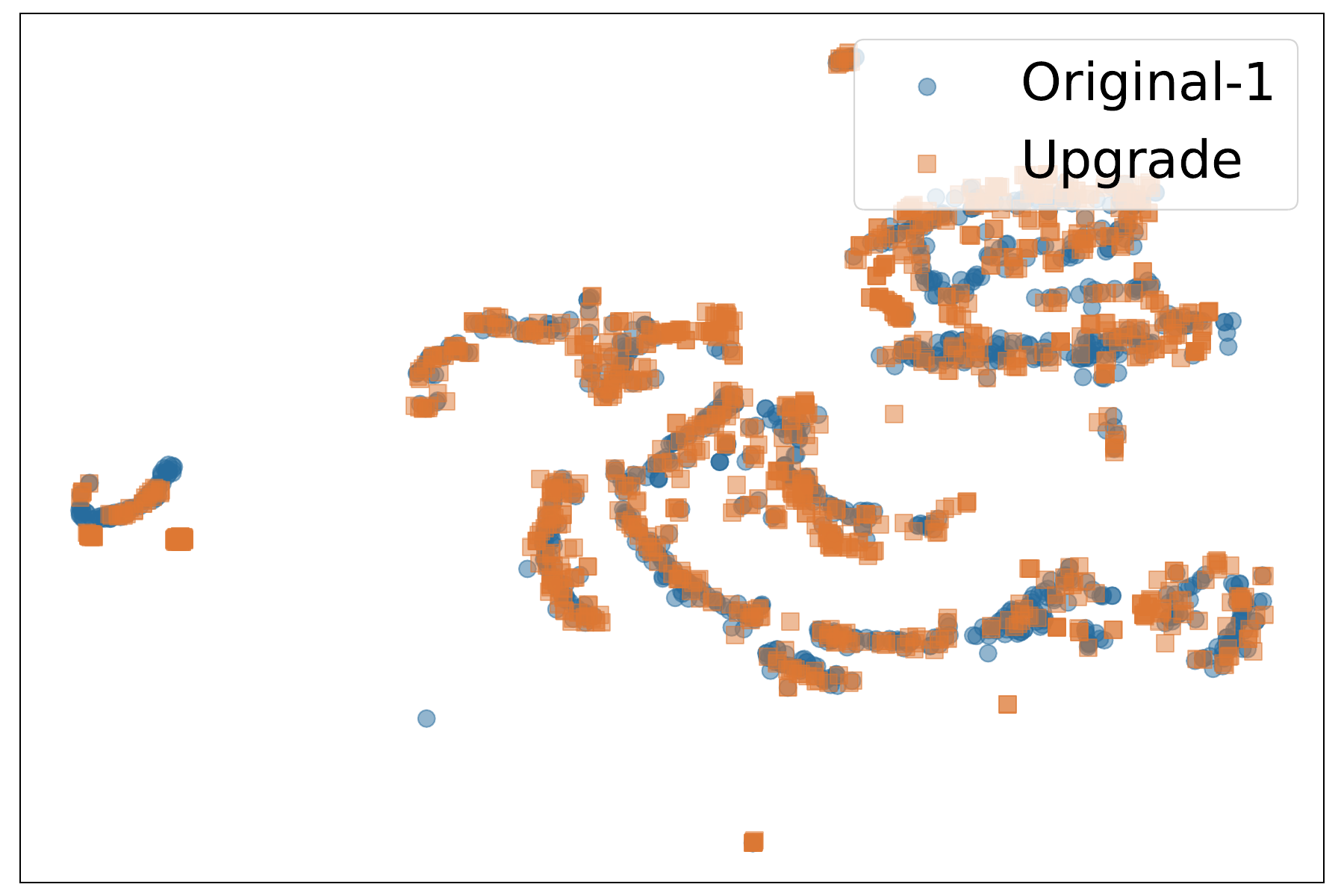}
	}
	\hfill
	\subfloat[SWaT-Mix]{
		\includegraphics[width=0.235\linewidth]{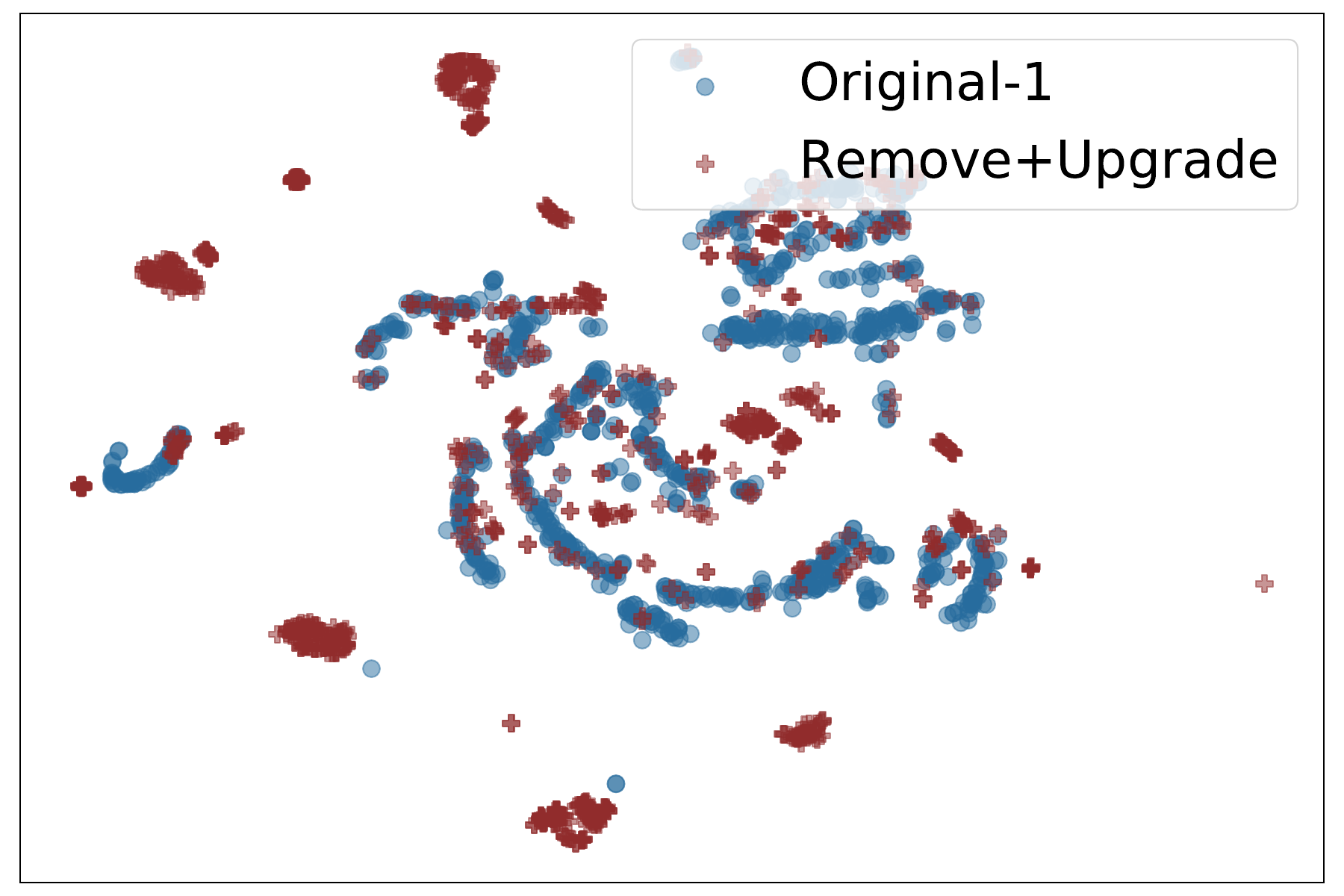}
	}

        \centering
        \scriptsize
	\hfill
	\subfloat[WADI-Original]{
		\includegraphics[width=0.235\linewidth]{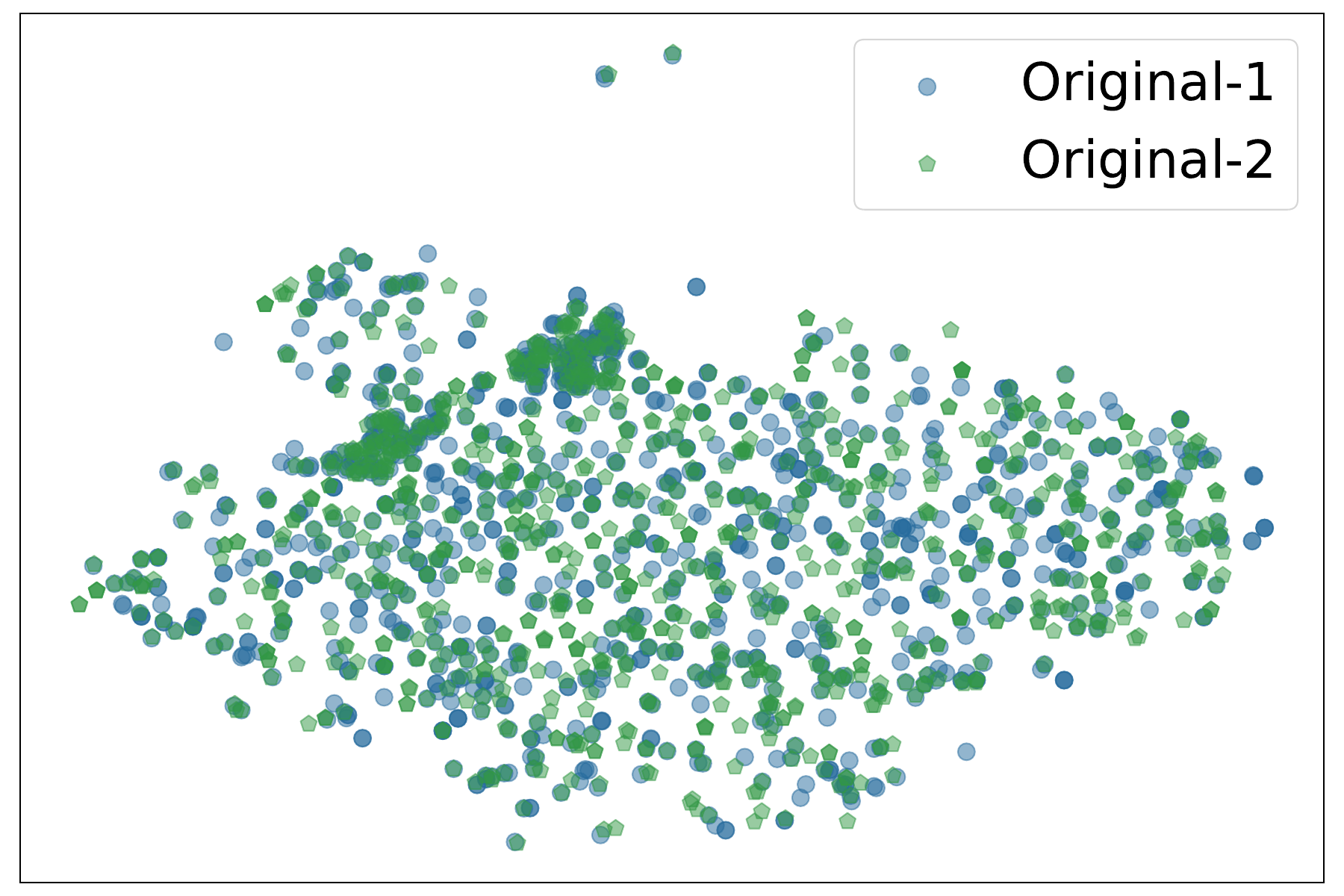}
	}
	\hfill
	\subfloat[WADI-Remove]{
		\includegraphics[width=0.235\linewidth]{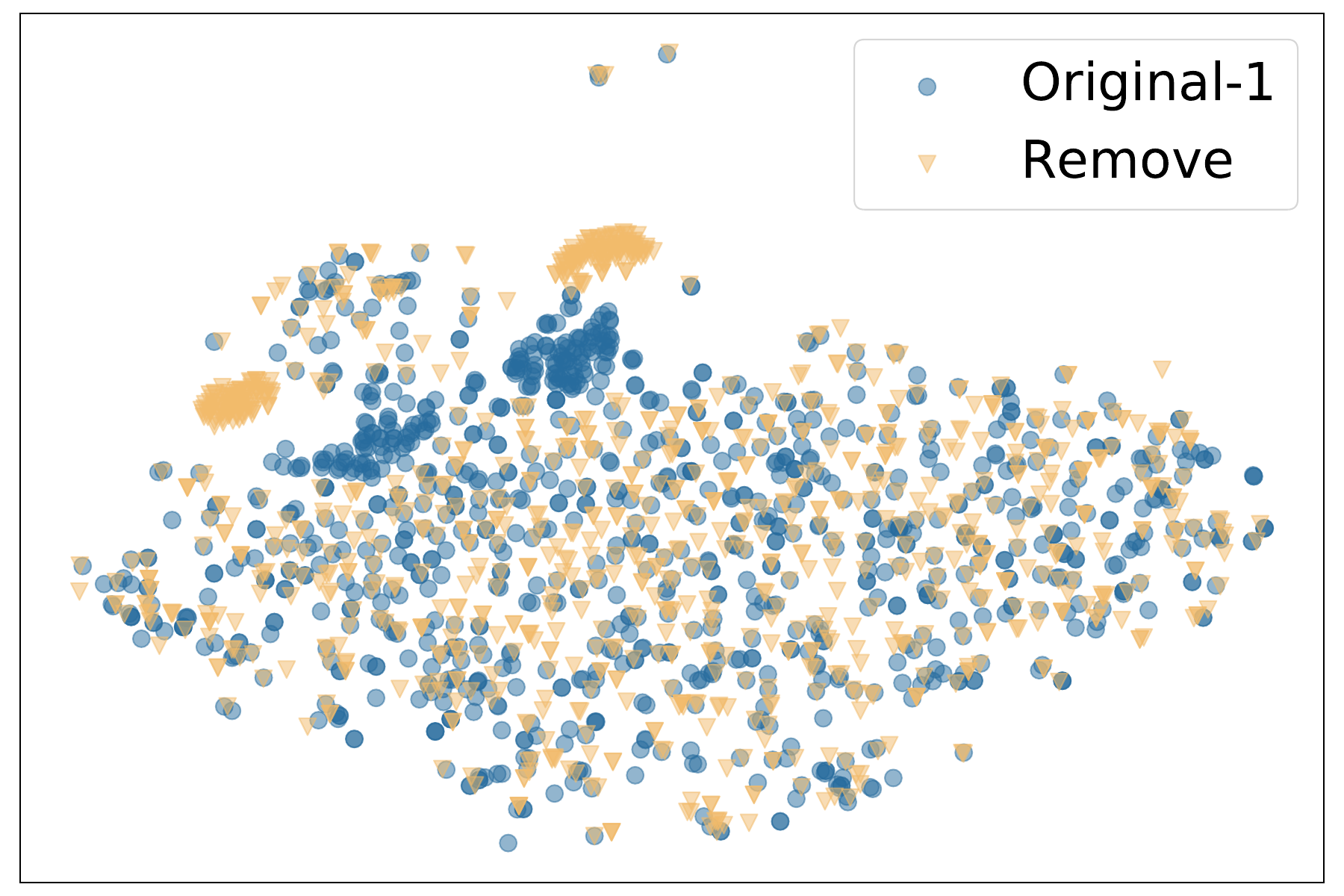}
	}
 	\hfill
	\subfloat[WADI-Upgrade]{
		\includegraphics[width=0.235\linewidth]{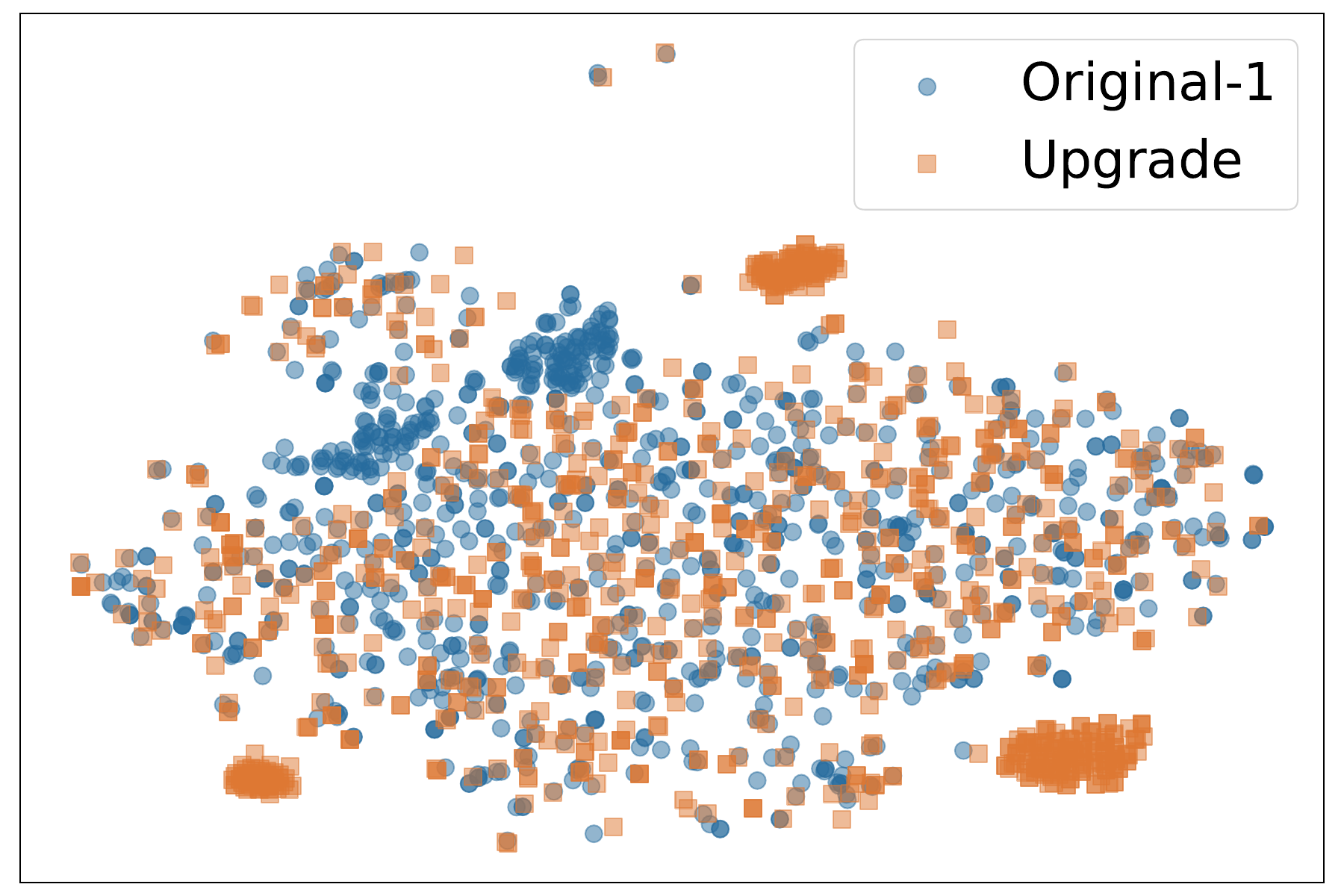}
	}
	\hfill
	\subfloat[WADI-Mix]{
		\includegraphics[width=0.235\linewidth]{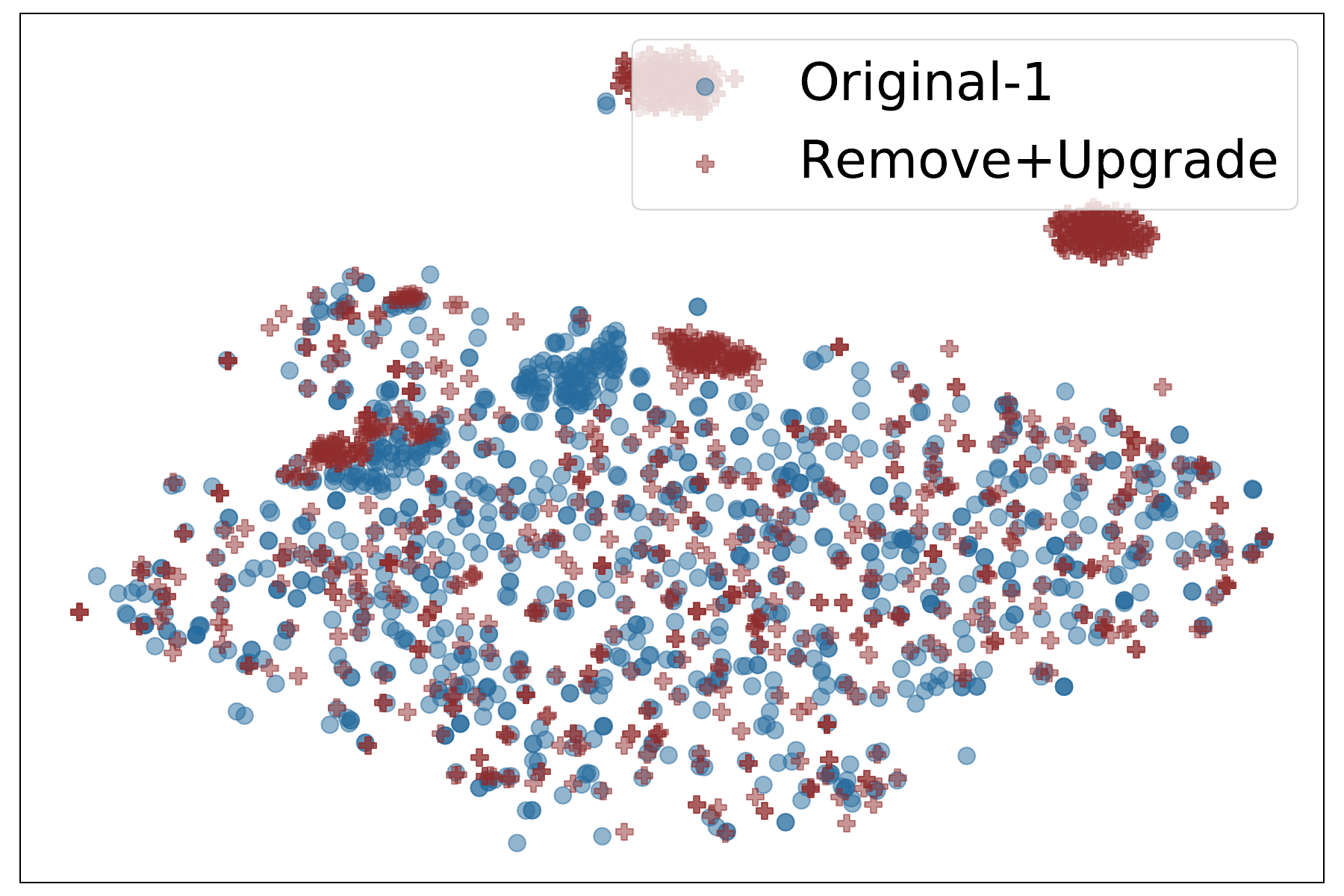}
	}
 
\caption{Intuitive demonstration of distribution shift from PUMP, SWaT, and WADI.}
\label{figure3}
\end{figure*}

During the operation of a CPS, both sensor and actuator devices inevitably dynamically change in number and performance. In particular, Sensors show improved adaptability, sensitivity, and resolution, while actuators demonstrate enhanced speed, power, torque, and energy efficiency. Users often replace older devices with superior ones offering enhanced performance. Meanwhile, Business requirements, such as modifications in industrial production lines, also influence the number of sensor and actuator devices.

Device changes directly impact the time series data distribution in the form of covariate shifts. To investigate this phenomenon, we take the open-source and widely adopted PUMP, SWaT, and WADI datasets as examples for data distribution visualization. Specifically,  to simulate CPS component evolution, we take the following three operations on the dataset: i) Remove: randomly removing five devices and their corresponding data to simulate the device reduction scenario; ii) Upgrade: randomly adjusting the values of five sensors and actuators uniformly within the range of -5\% to +5\% to simulate the devices performance upgrades; iii) Mix (Remove+Upgrade): Combining removal and adjustment operations to fully replicate CPS evolution complexities.

Compared to the original dataset, the simulated evolved dataset exhibits significant distribution gaps. Figures \ref{figure3} illustrate shifts in data distributions for PUMP (Remove and Mix), SWaT (Remove and Mix), and WADI (Upgrade and Mix). This gap poses a serious challenge to the generalization ability of models trained on historical data, leading to a degradation of their performance in the face of evolving data. Incrementally updating the model can help bridge this gap. However, it's crucial to note that both normal and anomalous samples inevitably rely on precious manual labeling. The scarcity of anomalies has prompted the existing ADCPS methods to be implemented based on unsupervised methods. CPS evolution aggravates this challenge, as we not only have difficulty in obtaining anomaly labels from evolving data, but we also have difficulty in obtaining large normal samples for frequent training of models employing continuous manual labeling.

\section{Methodology}
\label{sec:3}

\subsection{Overview}
\label{sec:3.1}

\begin{figure*}
\centering
\includegraphics[width=0.98\linewidth]{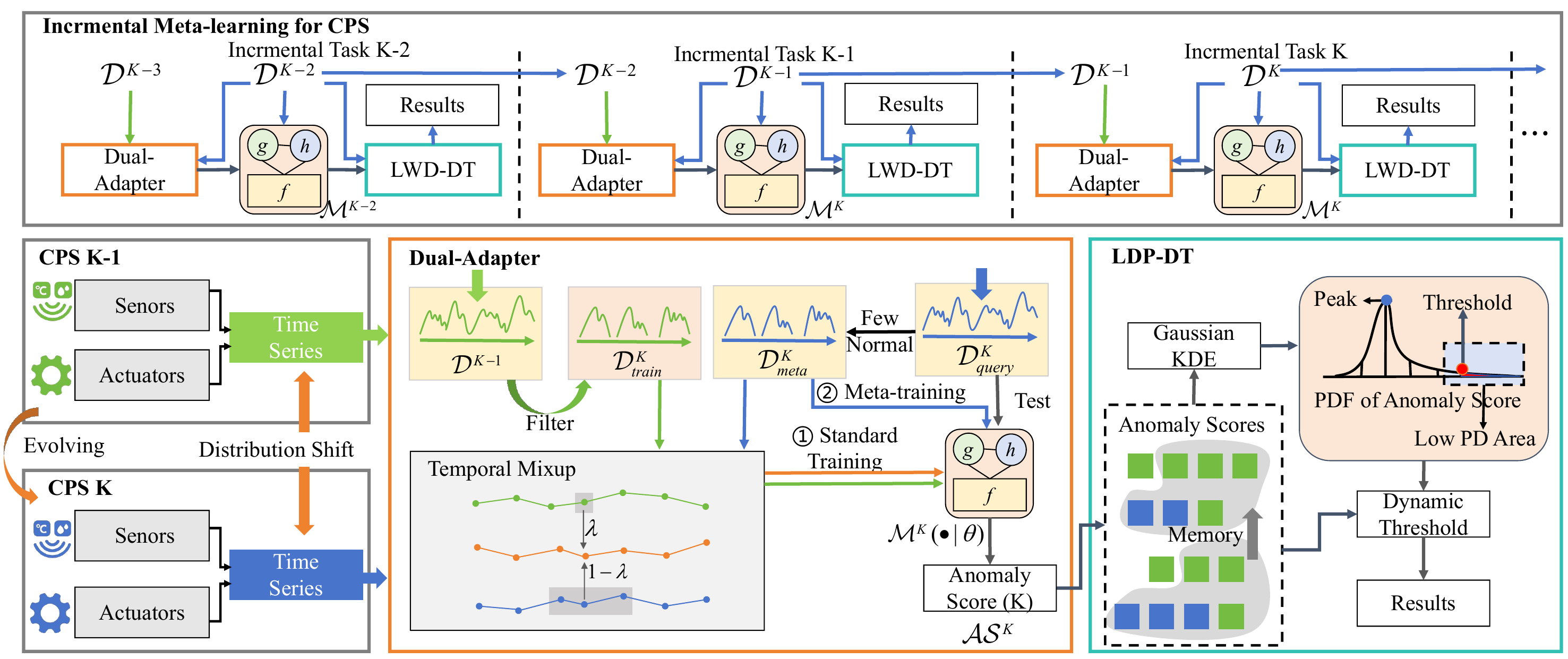}
\caption{Overview of iADCPS approach, which consists of two components: the Dual-Adapter, which achieves dual adaptation of the model $\mathcal{M}^{K}(\cdot|{\theta})$ through Temporal Mixup and incremental one-class meta-learning; and the LWD-DT, which dynamically adjusts thresholds based on low-density point without relying on the labels. } 
\label{framework}
\end{figure*}

In this paper, we propose iADCPS, an incremental meta-learning approach for detecting anomalies in evolving CPS. iADCPS adapts to novel time-series patterns using limited normal samples from the evolving CPS, focusing on both data and model levels and incorporates a dynamic thresholding algorithm based on low-density points for frequent model updates.  The methodology comprises two key components: the Dual-Adapter and LDP-DT.

\textbf{Dual-Adapter:} To achieve data alignment and model generalization with limited normal samples during incremental training, we combine the Mixup-based data augmentation algorithm and one-class meta-learning methods to propose a dual generalization strategy. Specifically, as shown in Figure~\ref{framework}, we first redesign the temporal Mixup algorithm \cite{cotmix} to generate mixed samples $\mathcal{D}^{K}_{mix}$ by mixing normal samples $\mathcal{D}^{K}_{train}$ filtered from the historical system time-series $\mathcal{D}^{K-1}$ with a few normal samples $\mathcal{D}^{K}_{meta}$ from the evolved system time-series $\mathcal{D}^{K}$, thus achieving the alignment at the data level. Subsequently, we implement incremental training based on one-class meta-learning using $\mathcal{D}^{K}_{train}$, $\mathcal{D}^{K}_{mix}$, and $\mathcal{D}^{K}_{meta}$ in two stages. Firstly, standard deep learning training is carried out using $\mathcal{D}^{K}_{train}$ and $\mathcal{D}^{K}_{mix}$ to ensure that the detection model can effectively extract features of the hidden state. Secondly, $\mathcal{D}^{K}_{meta}$ is then used to fine-tune the model utilizing meta-training so that the final model $\mathcal{M}^{K}(\cdot|{\theta})$ can be fully generalized to the evolved data. Finally, the evolved sample $\mathcal{D}^{K}_{query}$ is tested as a query set and the anomaly scores $\mathcal{AS}^{K}$ of the samples are output (see Section~\ref{sec:3.2} for details).

\textbf{LDP-DT:} To eliminate the dependence on anomaly labels and adaptively adjust the thresholds along with the incremental updates of the model, we designed LDP-DT. Initially, we employed Gaussian kernel density estimation (Gaussian-KDE) to construct the probability density function (PDF) of the sample anomaly scores $\mathcal{AS}^{K}$ (the number is limited by Memory for efficiency) without the need for prior knowledge. By identifying a low-density point near the x-axis along the PDF curve from the peak towards the right (reflecting increasing anomaly likelihood), we set the anomaly score at this point as the threshold. Samples surpassing this threshold are classified as abnormal. (see Section~\ref{sec:4} for details). 

For the K-th incremental task, we first use the Dual-Adapter to mix the normal samples $\mathcal{D}^{K}_{train}$ obtained from the historical time series with a few normal samples $\mathcal{D}^{K}_{meta}$ from the evolving time series to obtain the mixed dataset $\mathcal{D}^{K}_{mix}$. Then, we use $\mathcal{D}^{K}_{train}$ and $\mathcal{D}^{K}_{mix}$ as the merge set to train the end-to-end SSM $\mathcal{M}^{K-1}(\cdot|{\theta})$ (the structure of the SSM follows NSIBF \cite{nsibf}) from the previous task K-1 to obtain the model $\mathcal{M}^{K}(\cdot|\tilde{\theta})$. Subsequently, using $\mathcal{D}^{K}_{meta}$ as the support set for the meta-task and the test data as the query set $\mathcal{D}^{K}_{query}$, we perform meta-training of model $\mathcal{M}^{K}(\cdot|\tilde{\theta})$ to obtain a detection model $\mathcal{M}^{K}(\cdot|{\theta})$ with stronger generalization ability and calculate the anomaly scores of the test samples. Finally, we use the LDP-DT method to determine a suitable threshold for the evolving data of this task, and according to this threshold, we determine whether there are abnormal samples in testing data. 

\subsection{Dual-Adapter for Evolving CPS Anomaly Detection}
\label{sec:3.2}
To tackle the distribution shift challenge posed by CPS evolution, two important directions can be followed, i.e., i) reducing the distribution gap between evolved and training data, and ii) enhancing the generalization ability of the model to evolved data. Therefore, our incremental training framework addresses the challenge at both data and model levels.
 
\textbf{Temporal Mixup-based Data Adapter:} The Mixup algorithm generates mixed samples which feature space is closer to the evolving time series patterns by mixing different samples, which helps the model to learn a more generalized feature representation, and thus better adapt to the distribution of the evolved data. However, Existing Mixup strategies struggle with the continuous temporal data generated by sensors and actuators, compounded by the scarcity of labels in evolving systems. Drawing inspiration from previous work \cite{cotmix}, we devise a temporal Mixup strategy for mixing historical data with a few evolving data, thus avoiding the reliance on labels for frequent unsupervised incremental training. Since averaging time steps has the advantage of eliminating short-term fluctuations and reducing the effect of extreme values, we learn the temporal information in the evolving data by aggregating the \textit{N/2} forward and backward time steps of the evolving samples and mixing these time steps with one of the time steps in the historical samples, as shown in Figure~\ref{framework}. Specifically, each training data time step is combined with the average of \textit{N}time steps (including \textit{N/2} backward time steps and \textit{N/2} forward time steps) of the evolved data, where the proportion of the history samples is $\lambda$ and the proportion of the evolving sample is 1 - $\lambda$. Formally, for the \textit{i}-th history sample ${S}_{i}^{train}\in{D}^{K}_{train}$ and an evolving sample ${S}_{i}^{meta}\in{D}^{K}_{meta}$, the resulting mixed samples ${S}_{i}^{mix}\in{D}^{K}_{mix}$  can be formulated as:

\begin{align}
\label{E1}
{S}_{i,d}^{mix}={\lambda}{S}_{i,d}^{train}+(1-{\lambda})\frac{1}{{N}}\sum_{d-N/2}^{d+N/2}{S}_{i,d}^{meta}
\end{align}

where \textit{d} is the feature, \textit{N} is the mixed window length. Unlike~\cite{cotmix}, due to the small number of evolving normal samples, the generated mixing samples are more biased towards evolving features to get a better generalization of the model. We set the mixing rate $\lambda$ to be less than 0.5 so that the evolving samples dominate the mixing. Moreover, since we only utilize normal samples from the historical data and a few normal samples from the evolving data, the mixed samples are all labeled as normal.

\textbf{Meta-Learning-based Model Adapter:}
When oriented towards training data with limited labels,  meta-learning can more effectively generalize the model to unseen data. Therefore, in each incremental task, we use a training approach based on One-Class Meta-learning, which is divided into two phases to train the model. 

In the first phase, for the K-th incremental task, we utilize the historical samples $\mathcal{D}^{K}_{train}$, the mixed samples $\mathcal{D}^{K}_{mix}$ and evolving samples $\mathcal{D}^{K}_{meta}$ to form the merged training set $\mathcal{D}^{K}_{merge}$. The current weights of the model are set to $\mathcal{M}^{K-1}(\cdot|{\theta})$, and then the model is standard trained using $\mathcal{M}^{K-1}(\cdot|{\theta})$ and $\mathcal{D}^{K}_{merge}$ to obtain the updated weights $\tilde{\theta}^{K}$ to ensure that the SSM can effectively extract the hidden states (as shown in Section~\ref{sec:2.1}). The optimization objective is to reduce the training loss $\mathcal{L}_{train}$, which is defined as

\begin{align}
\label{E2}
\mathcal{L}_{train}=\frac1{\mid \mathcal{D}_{merge}^K\mid}\sum_{(x,y)\in \mathcal{D}_{merge}^K}\left(\mathcal{M}(x,\theta^{K-1})-y\right)^2
\end{align}

In the second stage, we use a few evolved normal samples as the meta-training set $\mathcal{D}^{K}_{meta}$, and split $\mathcal{D}^{K}_{meta}$ into multiple support sets for fine-tuning, to enable the model quickly generalize to evolved time-series patterns. The loss function for the meta-training phase can be formulated as

\begin{align}
\label{E3}
\mathcal{L}_{meta}=\frac1{\mid \mathcal{D}_{meta}^K\mid}\sum_{S^T\in \mathcal{D}_{meta}^K}\frac1{\mid S^T\mid}\sum_{(x,y)\in S^T}\left(\mathcal{M}(x,\tilde{\theta}^K)-y\right)^2
\end{align}

The parameters of the final model can be formulated as:

\begin{align}
\label{E4}
\boldsymbol{\theta}^K=\boldsymbol{\tilde{\theta}}^K-\boldsymbol{\eta}_\theta\nabla\boldsymbol{L}_{meta}(\boldsymbol{D}_{meta}^K;\boldsymbol{\tilde{\theta}}^K)
\end{align}

where $\eta$ is the meta-learning rate of the detection model. We perform a multi-step episode to achieve gradient updating. Then, the model $\mathcal{M}^{K}(\cdot|{\theta})$ for the K-th task is deployed online to test evolving samples $\mathcal{D}^{K}_{query}$ and output the anomaly scores. This training process aligns with the meta-learning objective of MAML\cite{maml}, focusing on optimizing the base parameter $\theta^{K-1}$ to swiftly adapt to new tasks with minimal gradient updates on limited normal data. As CPS systems continue evolving, there is a high probability that past data does not contain patterns that will re-emerge in the future. Therefore, we aim to improve the generalization performance of the model without paying attention to the problem of "catastrophic forgetting" in incremental learning (IL). In the incremental train setting, we preserve only samples from the preceding task in memory to streamline sample storage demands.

\section{Low-Density Point-based Dynamic Thresholding}
\label{sec:4}

\subsection{Dynamic Threshold}
\label{sec:4.1}

The evolution of the CPS leads to a constantly changing anomaly pattern, thus requiring regular model training to accommodate distribution shifts. Nevertheless, the predefined static threshold falls short of supporting continuous updating of the model.

In a recent study ADA \cite{ada}, the optimal threshold is determined by the point where the probability density distribution curves of normal and abnormal losses intersect in log data (refer to Figure \ref{figure5}(a)). It is easy to understand that in a binary classification problem, the intersection of the two types of loss distributions usually means that the sum of the true positive rate (TPR) and the false positive rate (FPR) is minimized. ADA methods, however, rely on abnormal labels, posing a challenge in practical CPS contexts where defining effective supervision can be complex. Furthermore, ADA operates on a normal loss distribution, contrasting with the density-based ADCPS approach that yields challenging-to-estimate anomaly scores.

\subsection{Dynamic Thresholding based on Low-Density Point}
\label{sec:4.2}
Although the ADA method encounters challenges in adapting to evolving CPS environments,  however, the method provides us with a valuable idea that a certain point in the PDF of the model's output (both loss and abnormal scores) represents the optimal threshold.

\begin{figure}[!htb]
	\centering
	\scriptsize
	\subfloat[ADA]{
		\includegraphics[width=0.48\linewidth]{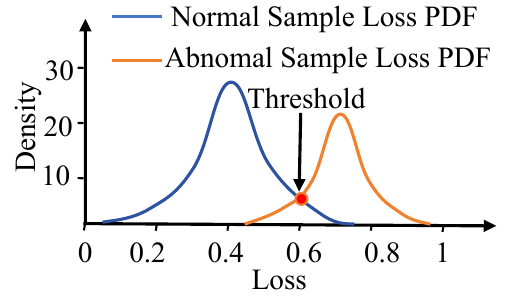}
	}
	\hfill
	\subfloat[LDP-DT]{
		\includegraphics[width=0.48\linewidth]{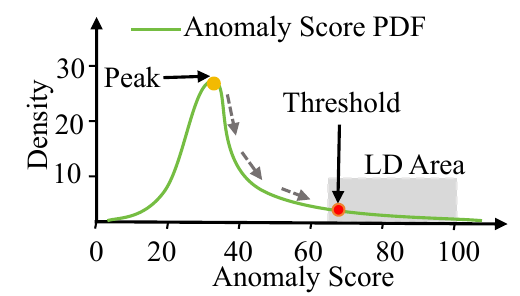}
	}
\caption{Comparison of ADA and LDP-DT.}
\label{figure5}
\end{figure}

Within ADCPS, the abnormal score probability density function (AS-PDF) characterizes the distribution density of samples, normal and abnormal alike, corresponding to different abnormal scores. Higher densities signify a greater number of samples linked to that anomaly score. Given the rarity of anomalies in CPS operations, it's commonly accepted that the system typically functions normally. Hence, it's reasonable to assume that normal sample anomaly scores tend to be lower and concentrated near the AS-PDF peak, while anomaly samples exhibit higher scores, dispersed within the low-density area (LD-Area) of the AS-PDF.

As shown in Figure \ref{figure5}, as we move away from the peak towards the right in the PDF, the anomaly score steadily increases while the density value gradually decreases, eventually nearing zero at the x-axis, creating a low-density area. It is easy to observe that the area signifies high anomaly scores for samples with a low likelihood of occurrence. Thus, we take the first point where the PDF tends to zero on the x-axis as the threshold point for anomaly detection, which is a reasonable choice based on the properties of probability distribution. Compared to fixed and dynamic thresholds, this approach can adapt to model updating without relying on anomaly labels. Notably, the LDP-DT method needs to be dynamically adjusted along with new samples. Therefore, in practice, to avoid too many anomalous scores to bring a large resource burden to the PDF calculation, we set up a Memory mechanism to retain only a limited number of anomalous scores. Under this mechanism, the anomaly scores of the old samples will be replaced by the anomaly scores of the new samples to ensure the computational efficiency of PDF.

\subsection{Threshold Calculation based on Gaussian Kernel Density Estimation}
\label{sec:4.3}
As discussed in Section ~\ref{sec:4.1}, the distribution of anomaly scores is not idealized normal, making it challenging to derive the PDF by parameter estimation. Gaussian Kernel Probability Density Estimation (Gaussian-KDE) offers a solution by approximating the PDF without requiring any prior assumptions. Therefore, we utilize Gaussian-KDE, a non-parametric estimation technique, to estimate the PDF of the sample anomaly scores.

For the anomaly scores set $x^{as}$, the basic idea of the Gaussian-KDE is to consider each query point as the center of a Gaussian kernel and to stack these kernels to form an estimate of the overall PDF. First, We can obtain the query points $x^{qp}$ to be used for evaluating the PDF by the following formula:

\begin{align}
\label{E5}
x^{qp}_{i}=x^{as}_{min}-3\sigma+\frac{i-1}{Z-1}[(x^{as}_{max}+3\sigma)-(x^{as}_{min}-3\sigma)]
\end{align}

$x^{qp}$ is an array containing \textit{Z} points uniformly distributed over the specified interval from min($x^{as}$)-3$\sigma$) to max($x^{as}$)+3$\sigma$). Here, min($x^{as}$) and max($x^{as}$) are the minimum and maximum values of the $x^{as}$, respectively, and $\sigma$ is the standard deviation of the $x^{as}$. Let $x^{as}_{min}$=min($x^{as}$), $x^{as}_{max}$=max($x^{as}$), $\sigma$=std($x^{as}$), and \textit{Z} is the number of query points. For each query point $x^{qp}_{i}$, its corresponding kernel function KER can be expressed as:

\begin{align}
\label{E6}
\mathrm{KER}(x^{qp}_i,x^{as}_i)=\frac{1}{\sqrt{2\pi}h}\exp(-\frac{(x^{qp}_i-x^{as}_i)^2}{2h^2})
\end{align}

For the entire anomaly score set $x^{as}$, the KDE of the $x^{as}$ at query points $x^{qp}_i$ can be expressed as a weighted sum of all the kernel functions, and the PDF are computed on $x^{as}$ can be the following formula:

\begin{align}
\label{E7}
\mathrm{KDE}(x^{qp}_i)=\frac1n\sum_{i=1}^n\frac1{\sqrt{2\pi}h}\exp(-\frac{(x^{qp}_i-x^{as}_i)^2}{2h^2})
\end{align}

\textit{h} is the variance of the Gaussian kernel, $h=(\frac{4}{3n})^{\frac{1}{5}}\sigma$. After calculating the KDEs corresponding to all the query points, we get the final PDF and can find the maximum point  (peak point) of the PDF using the following formula:

\begin{align}
\label{E9}
peak=\max_{i\in\{1,2,\ldots,Z\}}\mathrm{KDE}(x^{qp}_i)
\end{align}

Set a value of approximate precision $\delta$, from the peak point along the direction of the larger anomaly score to find, the first point found less than the approximate precision that is the low-density point, its corresponding anomaly score that is the threshold value. Then, a test sample whose output anomaly score is greater than the threshold value is an anomaly sample. 

\section{Qualitative Experiment}
In the qualitative experiments, we follow the NSIBF \cite{nsibf} and use sine wave simulated evolving data with variations in both amplitude and frequency to validate the generalization performances of the static method, the incremental training method, and the proposed method iADCPS. The implementation of the data simulation algorithm is described below:

\begin{align}
\label{E10}
t\%30=0\rightarrow u^{t}=9-u^{t} \\ \nonumber 
z^t=amp\bullet\sin(t/freq\bullet u^t)+\epsilon^t \\
x^{t}=2^{*}z^{t}+\varepsilon^{t} \nonumber 
\end{align}

where \textit{t}=$\{$1,2, ..., T$\}$, $x^{t}$ is the observed sensor measurements, $z^{t}$ is the hidden state of the system, and $u^{t}$ is the state of the actuator controlling the sinusoidal frequency. Given $u^{0}$=3, $\epsilon^{t} \in N(0,0.2^2)$ is the measurement noise and $\varepsilon^{t} \in N(0,0.2^2)$ is the process noise. First, We use the formula \ref{E10} to generate 10,000 time points to simulate the initial training data, with \textit{amps} and \textit{frep} set to 1. Meanwhile, we simulate five incremental training tasks with the amplitude set \textit{amps}=$\{$1.2, 1.4, 1.6, 1.8, 2.0$\}$ and the frequency set \textit{freqs}=$\{$2, 4, 6, 8, 10$\}$. 
To mimic the evolution of the CPS, we sequentially vary the amplitude \textit{amps} and frequency \textit{freqs} for each task, ranging from low to high. Each task is designed to produce 500 time points, simulating a limited set of normal samples as incremental training data using formula \ref{E10} and generating 2,000 time points for injecting anomalies as incremental test data. Specifically, in the incremental test data, 100 consecutive time points out of every 1000 time points are anomalous time points. At the abnormal time points, we set $\epsilon^{t} \in N(0,0.6^2)$.

To achieve time series-level anomaly detection, we employ a sliding-window strategy that combines sensor measurements from every 31 consecutive time points into a feature vector to ensure coverage of at least one complete cycle of actuator state changes. Based on the complete initial training dataset, we first obtain a pre-trained model based on standard training. Following this, the model is utilized in three distinct scenarios: static detection, incremental training, and the proposed iADCPS approach:

i) Static: In this mode, the pre-trained model is directly employed for detection without incorporating incremental learning strategies.

ii) Incremental Training (IT): the detection model undergoes continuous updates through incremental training data that evolves.

iii) iADCPS: this method involves incremental training facilitated by temporal mixup algorithms and meta-learning.

\begin{figure}[!htb]
	\centering
	\scriptsize
	\subfloat[Anomaly Score]{
		\includegraphics[width=0.48\linewidth]{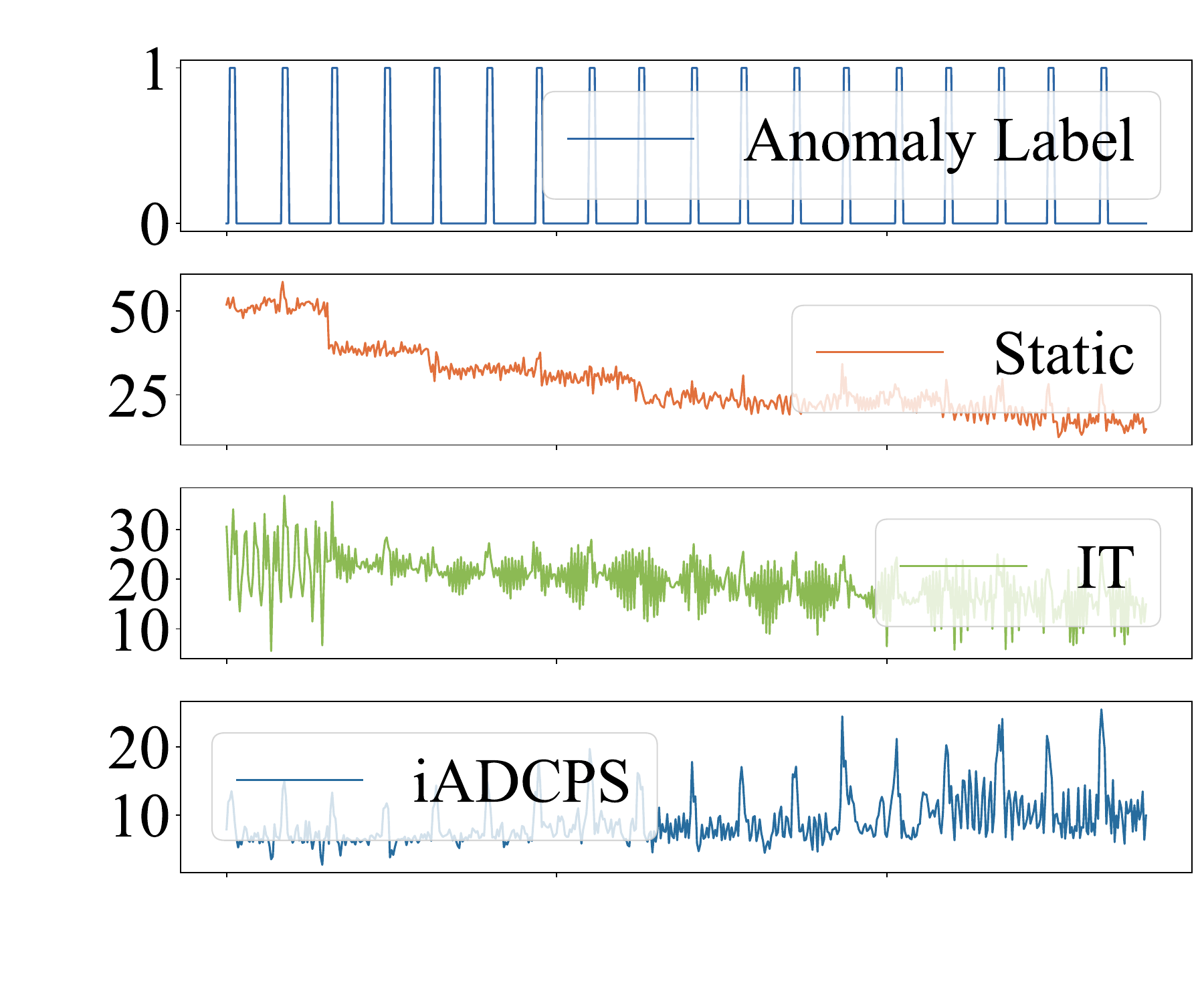}
	}
	\hfill
	\subfloat[ROC Curve]{
		\includegraphics[width=0.48\linewidth]{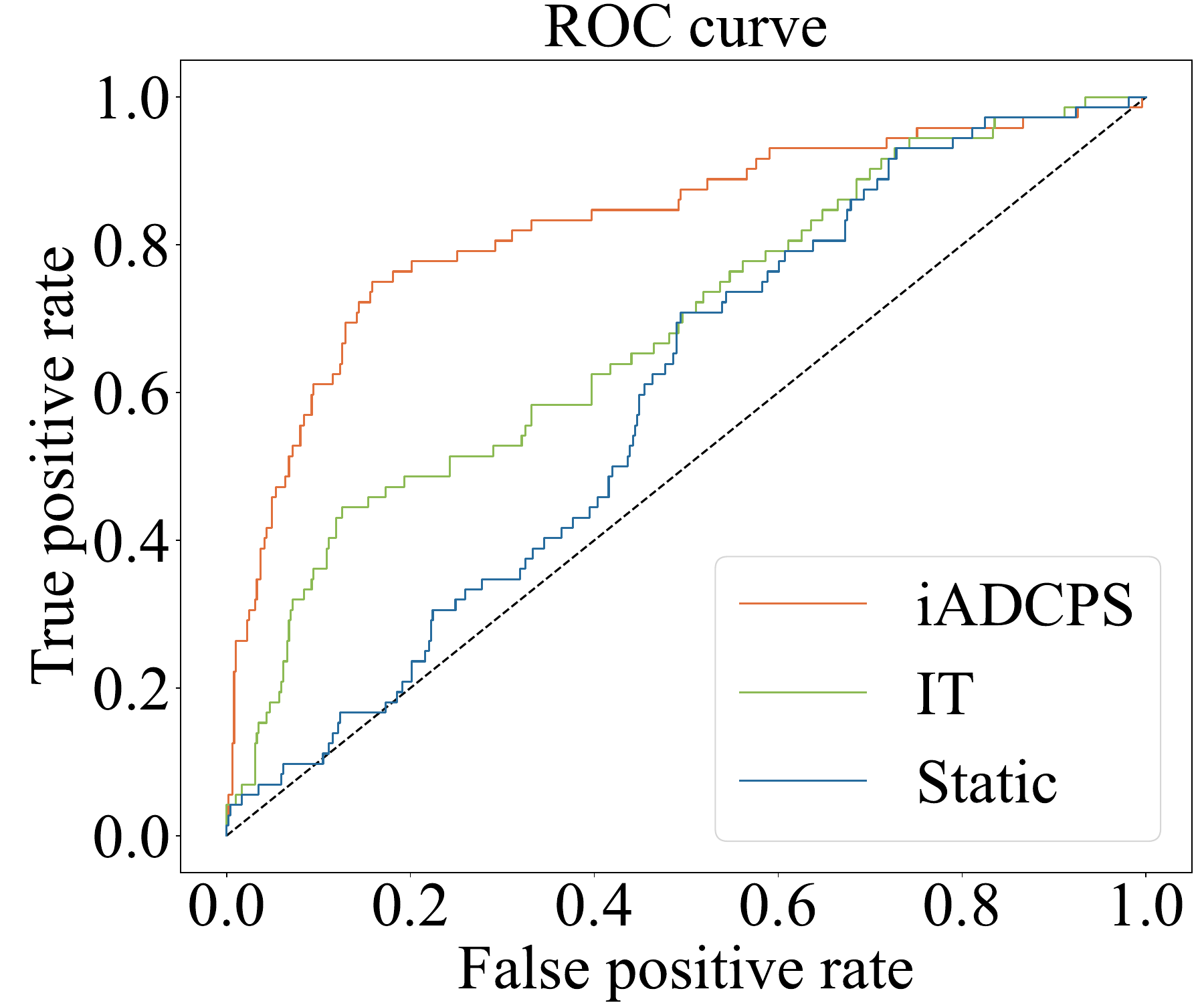}
	}
\caption{Results of qualitative experiment.}
\label{figure6}
\end{figure}

Figure~\ref{figure6} (a) illustrates the distribution of anomaly scores generated by the three methods, Static, IT, and iADCPS, while Figure~\ref{figure6} (b) provides their corresponding ROC curves for evaluating the anomaly detection performance. It can be observed that iADCPS with the incremental meta-learning approach exhibits higher sensitivity and accuracy in detecting injected anomalies, which is mainly attributed to its efficient generalization capability through the incremental meta-training mechanism.

\section{Quantitative Experiment}
\label{sec:6}
In this section, we conducted experiments on three real datasets PUMP, WADI, and SWaT, and compared them with nine SOTA ADCPS methods. We investigate the effectiveness and limitations of the proposed method by addressing the following four research questions:

RQ1: How does iADCPS compare with SOTA ADCPS methods on stable datasets?

RQ2: How does iADCPS compare with SOTA ADCPS methods on evolving datasets?

RQ3: What are the training efficiency and detection efficiency of iADCPS?

RQ4: What are the contributions of each component to iADCPS?

\subsection{Dataset}
PUMP dataset is a time-series dataset sourced from water treatment and distribution systems, spanning from April 1st, 2018 to September 1st, 2018, covering a total of five months. Data is collected at a frequency of once per minute, encompassing 52 sensor measurements. The pump status is categorized into three types: broken, recovering, and normal, with a total of seven instances of the broken state. This dataset can be utilized to investigate the detection capabilities of ADCPS methods for CPS system failures.

SWAT dataset originates from an operational testbed of a water treatment system in the real world, designed for producing filtered water. It comprises 51 sensor measurements and actuator status values. The dataset encompasses 11 consecutive days of operation, with 7 days representing normal conditions and 4 days simulating attack scenarios.

WADI dataset builds upon SWaT and is specifically designed for secure water distribution. It includes 123 sensor measurements and actuator status values. The dataset spans 14 days under normal conditions and 2 days under attack scenarios.

In the stable experimental scenario, we follow the settings in NSIBF to split the data into a training set and a test set without making any changes. In the evolved experimental scenario, we further split the test set into incremental task subsets, where each task subset contains 4,000 data points. We select 1,000 data points as the incremental training set for simulating the few normal samples in the evolving scenario, and the remaining 3000 samples as the incremental test set. To maintain the consistency of the evaluation, we use the same input settings \cite{nsibf} for methods after data pre-processing, i.e., the sliding window lengths of PUMP, WADI, and SWAT are set to 5, 12, and 12, respectively. To evaluate the detection effectiveness of the iADCPS method for evolving systems, we set up three evolving scenarios, including:

Remove: i.e., randomly delete data corresponding to 1-5 sensors and actuators in the incremental task subset to simulate device removal.

Upgrade: i.e., randomly select the data corresponding to 1-5 sensors and actuators in the task subset with ±5\% value floating to simulate the scenario of device upgrade.

Mix: i.e., the dual operations of removing and replacing devices are performed on the dataset at the same time to simulate complex system evolution scenarios.

\subsection{Baselines and Evaluation Criteria}
To validate the performance of iADCPS, we selected several SOTA anomaly detection methods as a baseline, including the static Isolation Forest (IF)~\cite{if2008}, LSTM-VAE~\cite{lstm_vae}, USAD~\cite{usad}, NSIBF~\cite{nsibf}, FuSAGNet~\cite{fusagnet}, FuGLAD~\cite{fuglad}, CTAD~\cite{ctad}, CutAddPaste~\cite{cutaddpaste} and the dynamic LSTM-NDT~\cite{lstm_ndt}. As a classical machine learning one classification method, IF achieves effective data isolation and anomaly identification by constructing a binary tree structure; AE uses an autoencoder with sparse hidden embedding for anomaly detection; LSTM-Pred uses an LSTM-based regressor as a prediction model; USAD amplifies anomalous reconstruction errors using two autoencoders within an adversarial training framework; and NSIBF utilizes a state-space model based on neural network models to achieve end-to-end anomaly detection; FuSAGNet integrates sparse autoencoders with graph neural networks, while FuGLAD further incorporates structured prior knowledge for graph structure learning; CTAD employs self-supervised learning for model training, and CutAddPaste integrates domain knowledge to enhance anomaly data augmentation. 

The essence of incremental CPS anomaly detection remains the binary categorization of samples and the test results are categorized into four scenarios:

TP: Correctly detects abnormal samples as anomalous.

FN: Incorrectly detects abnormal samples as normal

TN: Correctly detects normal samples as normal

FP: Incorrectly detects normal samples as abnormal.

To evaluate the effectiveness of the anomaly detection model in incremental scenarios, we use the F1-score as an evaluation metric. This metric is a comprehensive evaluation metric that combines Precision and Recall. Among them, Precision (PRE) is the proportion of all samples identified as strange anomalies. Recall (REC) is the proportion of samples that are correctly identified as anomalous out of all anomalous samples. The F1-score (F1) is the reconciled average of Precision and Recall, not the summed average.

\subsection{Implementation and Environment}
In our experiments, the anomaly detection model follows the state-space model structure of NSIBF (i.e., the three sub-network models of $\textit{f}_{net}$, $\textit{g}_{net}$, and $\textit{h}_{net}$ to ensure the robustness of the detection. We use Adam as the optimizer, set the learning rate to 0.00001, mixing rate
$\lambda$ to 0.2, number of the query point \textit{Z} to 1000, approximate precision $\delta$ to 0.05, epoch to 100, and incremental meta-training episode to 10. To complete the testing of the method, the experiments use the following hardware and software platforms: an Intel Xeon Gold 5218 CPU @ 2.30GHz, CPU core 16, 256GB RAM, a NVIDIA Tesla V100-PCIE 32GB x 2, CentOS Linux version 7.9, CUDA 11.7, cuDNN 7.6.5, Python 3.7.12 and TensorFlow 2.3.0.

\subsection{RQ1. Detection performance on the Stable Data}

\begin{figure*}[!htb]
	\centering
	\scriptsize
	\subfloat[PUMP]{
		\includegraphics[width=0.32\linewidth]{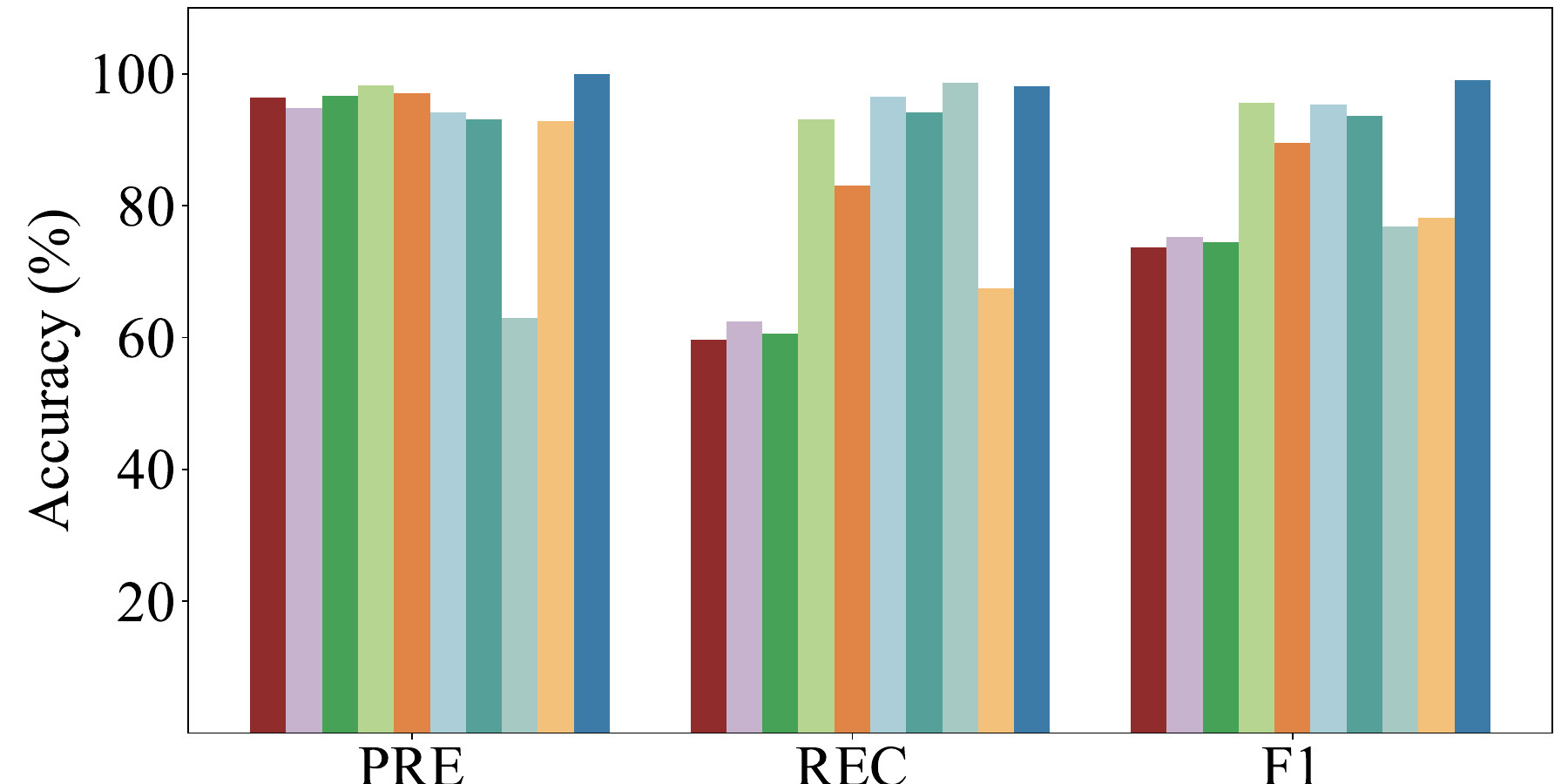}
	}
	\hfill
	\subfloat[SWaT]{
		\includegraphics[width=0.32\linewidth]{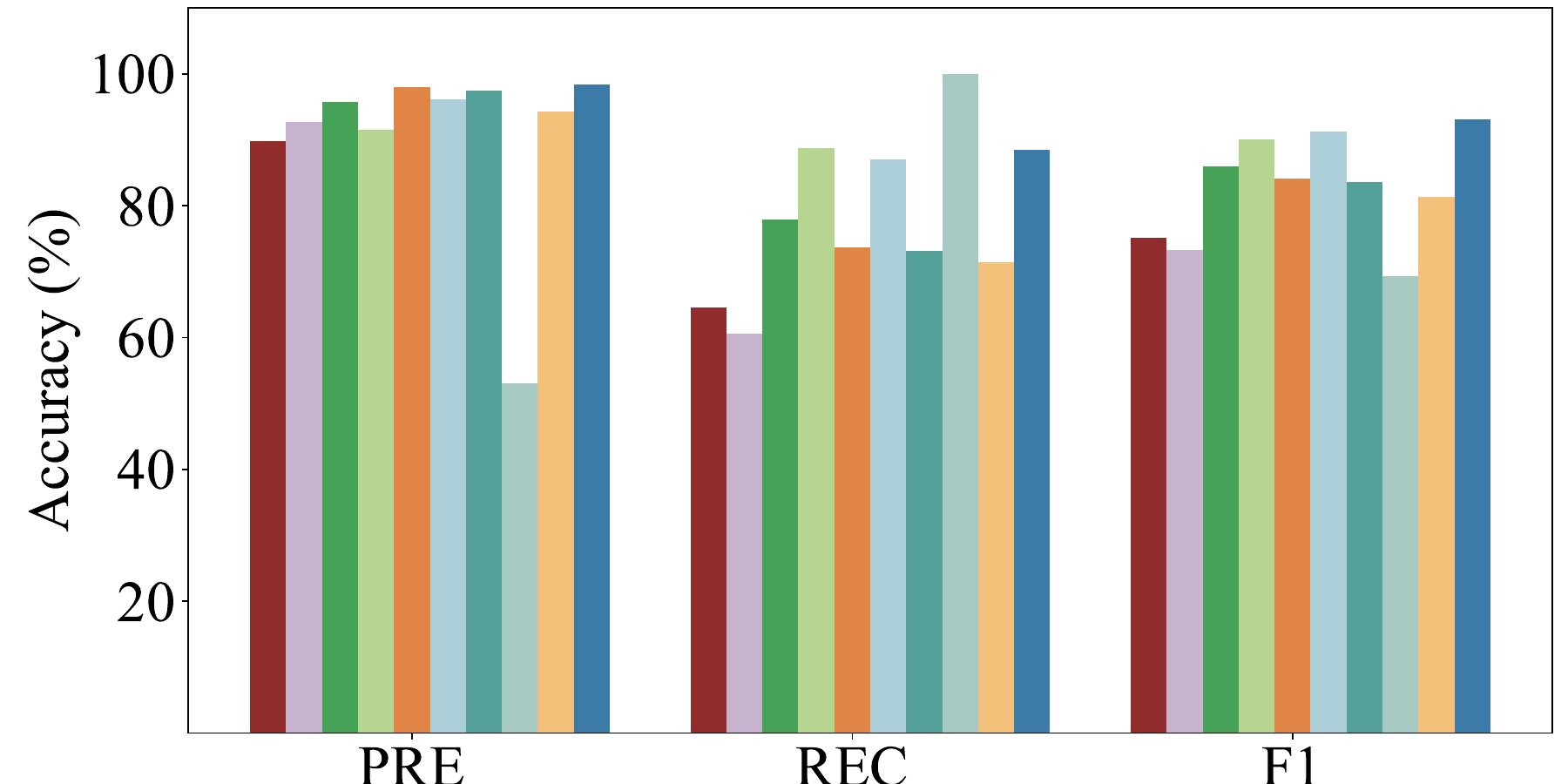}
	}
	\hfill
	\subfloat[WADI]{
		\includegraphics[width=0.32\linewidth]{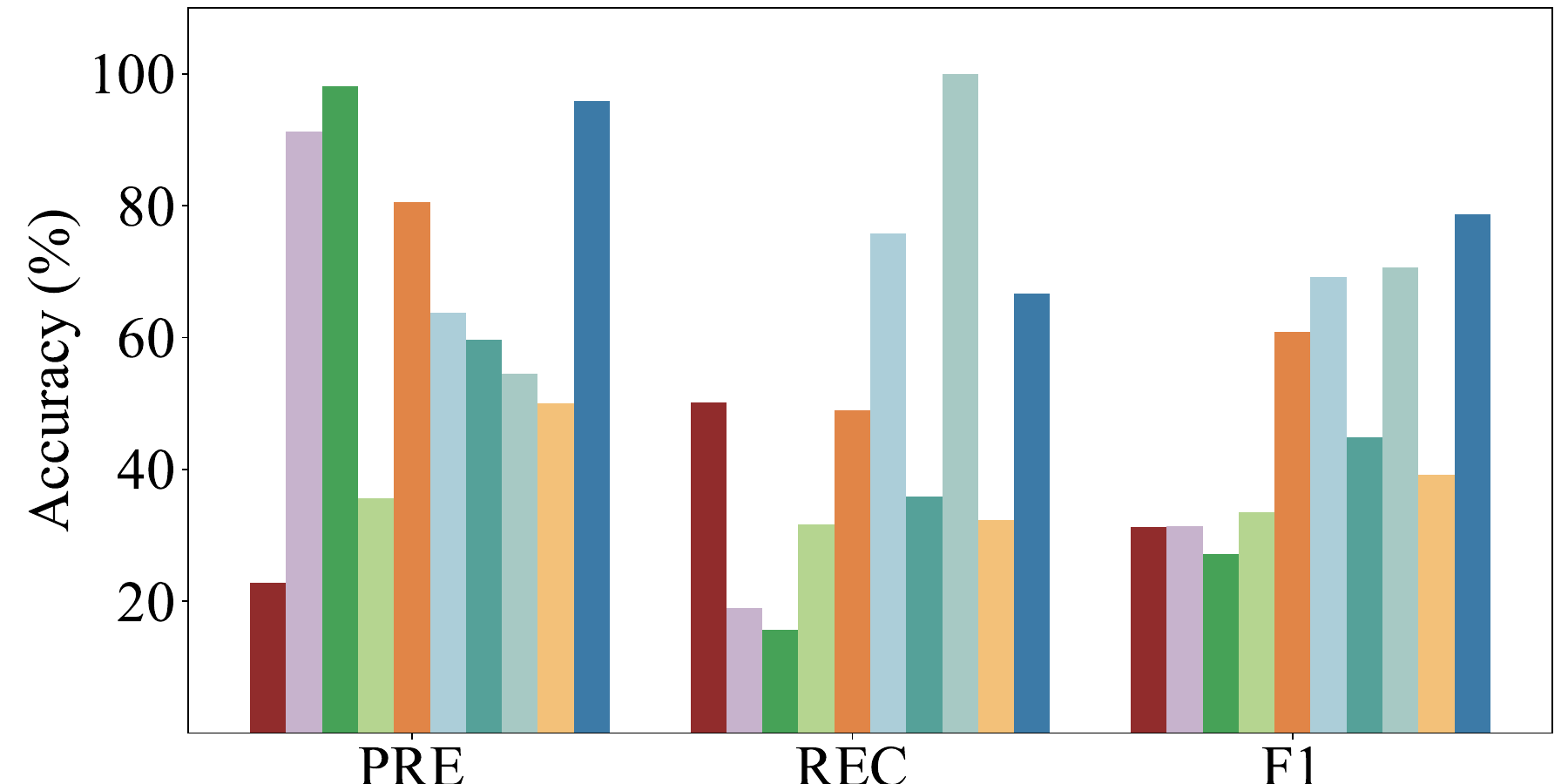}
	}
 
        \centering
        \scriptsize
	\subfloat{
		\includegraphics[width=0.6\linewidth]{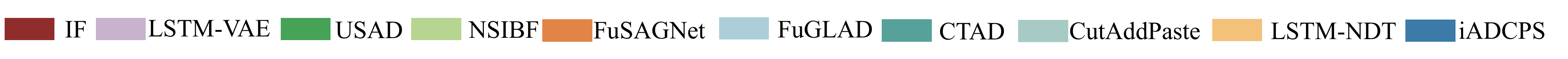}
	}
\caption{Result of the SOTA ADCPS approaches on stable CPS time series.}
\label{figure7}
\end{figure*}

\begin{table*}[ht]
\newcommand{\tabincell}[2]{\begin{tabular}{@{}#1@{}}#2\end{tabular}}
\centering
\begin{threeparttable}
\setlength\tabcolsep{7pt}
\caption{Results of the SOTA ADCPS approaches on Evolving time series}
\begin{tabular}{cccccccccccc}
\toprule  
\multirow{2}*{Scenarios} &\multirow{2}*{Types}&\multirow{2}*{Methods} & \multicolumn{3}{c}{PUMP} & \multicolumn{3}{c}{SWaT}& \multicolumn{3}{c}{WADI}\\
\cline{4-12}~ &~ &~ &PRE & REC & F1 & PRE & REC & F1& PRE & REC & F1\\
\midrule
\multirow{8}*{Remove}&\multirow{6}*{Static}&IF&96.1&	52.8&	68.2&	92.3&	57.5&	70.9&	16.6&	41.6&	23.7 \\
~&~&LSTM-Pred&84.1&	64.6&	73.1&	81.8&	59.5&	68.9&	79.3&	16.4&	27.2 \\
~&~&USAD&95.2&	55.7&	70.3&	93.1&	66.6&	77.7&	92.4&	11.2&	20 \\
~&~&NSIBF&98.4&	79.8&	88.1&	64.2&	100&	78.2&	34.4&	26.1&	29.7 \\
~&~&FuSAGNet&95.7&	74.5&	83.8&	94.9&	67.3&	78.8&	74.3&	43.5&	54.9 \\
~&~&FuGLAD&\textbf{88.6}	&\textbf{93.8}	&\textbf{91.1}&	\textbf{89.9}&	\textbf{83.9}&	\textbf{86.8}&	58.1&	70.1&	63.5 \\
~&~&CTAD&86.3&	92.6&	89.3&	97.2&	65.7&	78.4&	50.9&	33.0&	40.0 \\
~&~&CutAddPaste&56.8&	98.7&	72.1&	50.4&	94.9&	65.8&	\textbf{50.6}&	\textbf{100}&	\textbf{67.2} \\
\cline{2-12}
~&\multirow{2}*{Dynamic}&LSTM-NDT&89.3&	63.5&	74.2&	86.7&	67.2&	75.7&	42.2&	27.7&	33.4 \\
~&~&iADCPS&100&	98.1&	99.0&	92.3&	89.5&	90.9&	88.2&	64.7&	74.6 \\
~&~&Percentage Increase&+11.4\%&	+4.3\%&	+7.9\%&	+2.4\%&	+5.6\%&	+4.1\%&	+37.6\%&	-35.3\%&	+7.4\%\\
\bottomrule
\multirow{8}*{Upgrade}&\multirow{6}*{Static}&IF&91.1&	58.0&	70.9&	92.6&	60.0&	72.8&	21.6&	49.0&	30.0 \\
~&~&LSTM-Pred&91.5&	62.4&	74.2&	92.1&	58.2&	71.3&	81.1&	18.4&	30.0 \\
~&~&USAD&95.0&	58.1&	72.1&	97.8&	71.5&	82.6&	87.5&	15.0&	25.6 \\
~&~&NSIBF&\textbf{95.5}&	\textbf{93.1}&	\textbf{94.3}&	\textbf{91.3}&	\textbf{85.8}&	\textbf{88.5}&	34.4&	30.6&	32.4\\
~&~&FuSAGNet&92.9&	81.4&	86.8&	97.5&	72.5&	83.2&	75.6&	48.3&	58.9 \\
~&~&FuGLAD&91.9	&95.6&	93.7&	92.8&	82.7&	87.5&	63.3&	73.5&	68.0 \\
~&~&CTAD&92.0&	93.3&	92.6&	96.7&	71.4&	82.1&	57.9&	34.1&	42.9 \\
~&~&CutAddPaste&62.4&	96.9&	75.9&	50.0&	100&	66.7&	\textbf{52.8}&	\textbf{100}&	\textbf{69.1} \\
\cline{2-12}
~&\multirow{2}*{Dynamic}&LSTM-NDT&90.5&	66.6&	76.7&	90.7&	70.3&	79.2&	44.0&	34.0&	38.4\\
~&~&iADCPS&100&	98.1&	99.0&	94.6&	85.3&	89.7&	92.7&	65.2&	76.6 \\
~&~&Percentage Increase&+4.5\%&	+5.0\%&	+4.7\%&	+3.3\%&	-0.5\%&	+1.2\%&	+39.9\%&	-34.8\%&	+7.5\%\\
\bottomrule
\multirow{8}*{Mix}&\multirow{6}*{Static}&IF&96.0&	50.5&	66.2&	77.1&	58.2&	66.3&	15.7&	43.6&	23.1 \\
~&~&LSTM-Pred&84.0&	64.2&	72.8&	81.6&	58.2&	67.9&	77.4&	15.0&	25.1 \\
~&~&USAD&87.0&	56.8&	68.7&	78.7&	66.6&	72.1&	92.4&	11.1&	19.8 \\
~&~&NSIBF&94.6&	82.0&	87.9&	63.6&	100&	77.8&	34.3&	25.9&	29.5\\
~&~&FuSAGNet&94.6&	73.3&	82.6&	82.9&	64.3&	72.4&	67.6&	44.2&	53.5\\
~&~&FuGLAD&\textbf{88.6}&	\textbf{92.6}&	\textbf{90.6}&	\textbf{84.3}&	\textbf{79.5}&	\textbf{81.8}&	56.9&	67.8&	61.9 \\
~&~&CTAD&84.9&	91.5&	88.1&	96.7&	62.6&	76.0&	49.0&	32.3&	38.9 \\
~&~&CutAddPaste&55.2&	98.6&	70.8&	48.1&	94.2&	63.7&	\textbf{50.5}&	\textbf{100}	&\textbf{67.1} \\
\cline{2-12}
~&\multirow{2}*{Dynamic}&LSTM-NDT&91.0&	59.2&	71.7&	84.2&	65.3	&73.6&	42.3&	27.8&	33.6\\
~&~&iADCPS&99.8&	99.8&	99.8&	96.1&	72.2&	82.5&	86.1&	61.6&	71.8 \\
~&~&Percentage Increase&+11.2\%&	+7.2\%&	+9.2\%&	+11.8\%&	-7.3\%&	+0.7\%&	+35.6\%&	-38.4\%&	+4.7\%\\
\bottomrule
\end{tabular}
\label{table1}
\end{threeparttable}
\end{table*}

We first conducted experiments on the stable dataset, and the results of the iADCPS method compared with other methods are shown in Figure \ref{figure7}. Compared to existing ADCPS methods, iADCPS achieves the best results on each dataset. On the PUMP dataset, the precision, recall, and F1 of iADCPS are 100\%, 98.1\%, and 99.0\%, respectively. Our proposed method successfully detects all anomalies while generating only very few false alarms. On the SWAT and WADI datasets, iADCPS achieves F1 of 91.1\% and 78.7\%, respectively, again outperforming existing SOTA methods.

Notably, as a member of the static detection methods, the NSIBF method followed by iADCPS demonstrates a significant advantage in detection performance. On PUMP and SWaT, the F1 of NSIBF exceeds 90\%, achieving SOTA results among static methods. This demonstrates that NSIBF based on adaptive SSM and Bayesian Filtering is more suitable for constructing ADCPS models. However, on the complex WADI dataset, NSIBF still exhibits limitations and underperforms compared to CutAddPaste. Comprehensively, both approaches demonstrate significantly inferior detection performance relative to the proposed iADCPS.

\subsection{RQ2. Detection performance on the Evolving Data}
It is easy to find that the proposed iADCPS method achieves optimal results on the PUMP, WADI, and SWAT datasets. Especially on the PUMP dataset, the F1 all exceed 99\%, achieving accurate capture of anomalous events. It must be acknowledged that the evolution of the data also has an impact on iADCPS. This is particularly evident in the SWAT dataset. For example, on data that has not evolved, the PRE, REC, and F1 of iADCPS on the SWAT dataset are 98.4\%, 88.4\%, and 93.1\%, respectively. However, in the CPS mixed scenario, although iADCPS still leads, its PRE, REC, and F1 drop to  96.1\%, 72.2\%, and 82.5\%, respectively.

Notably, the static detection methods are clearly inadequate when dealing with evolving data. For example, in the stable scenario, the CutAddPaste method has an F1 of 70.6\% on the WADI dataset. However, under the mixed scenario, the F1 of the NSIBF method on WADI is reduced to 67.1\%, which is completely unable to achieve effective anomaly detection. It is worth noting that the detection performance of the dynamic threshold-based detection method, LSTM-NDT, is not outstanding, and even lower than the static detection method on most scenarios and datasets. After analysis, this is mainly because only the dynamic adjustment of thresholds is concerned in the LSTM-NDT method without timely updating of model parameters, which results in the LSTM-based detection model not being able to effectively generalize to evolved data.

\subsection{RQ3. Detection efficiency on Evolving Data}
\begin{figure}[!htb]
	\centering
	\scriptsize
	\subfloat[PUMP Training Time]{
		\includegraphics[width=0.48\linewidth]{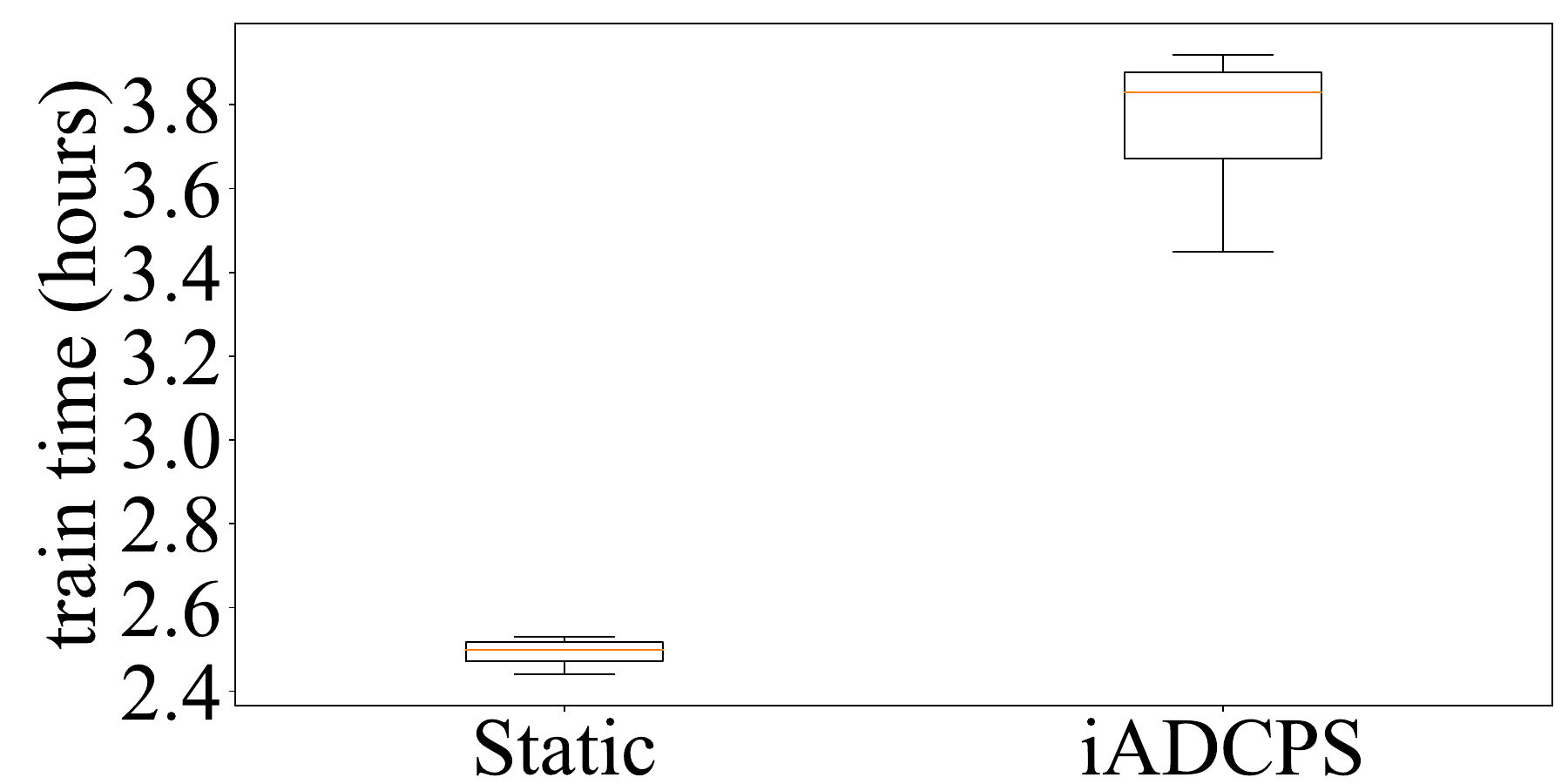}
	}
	\hfill
	\subfloat[PUMP Testing Time ]{
		\includegraphics[width=0.48\linewidth]{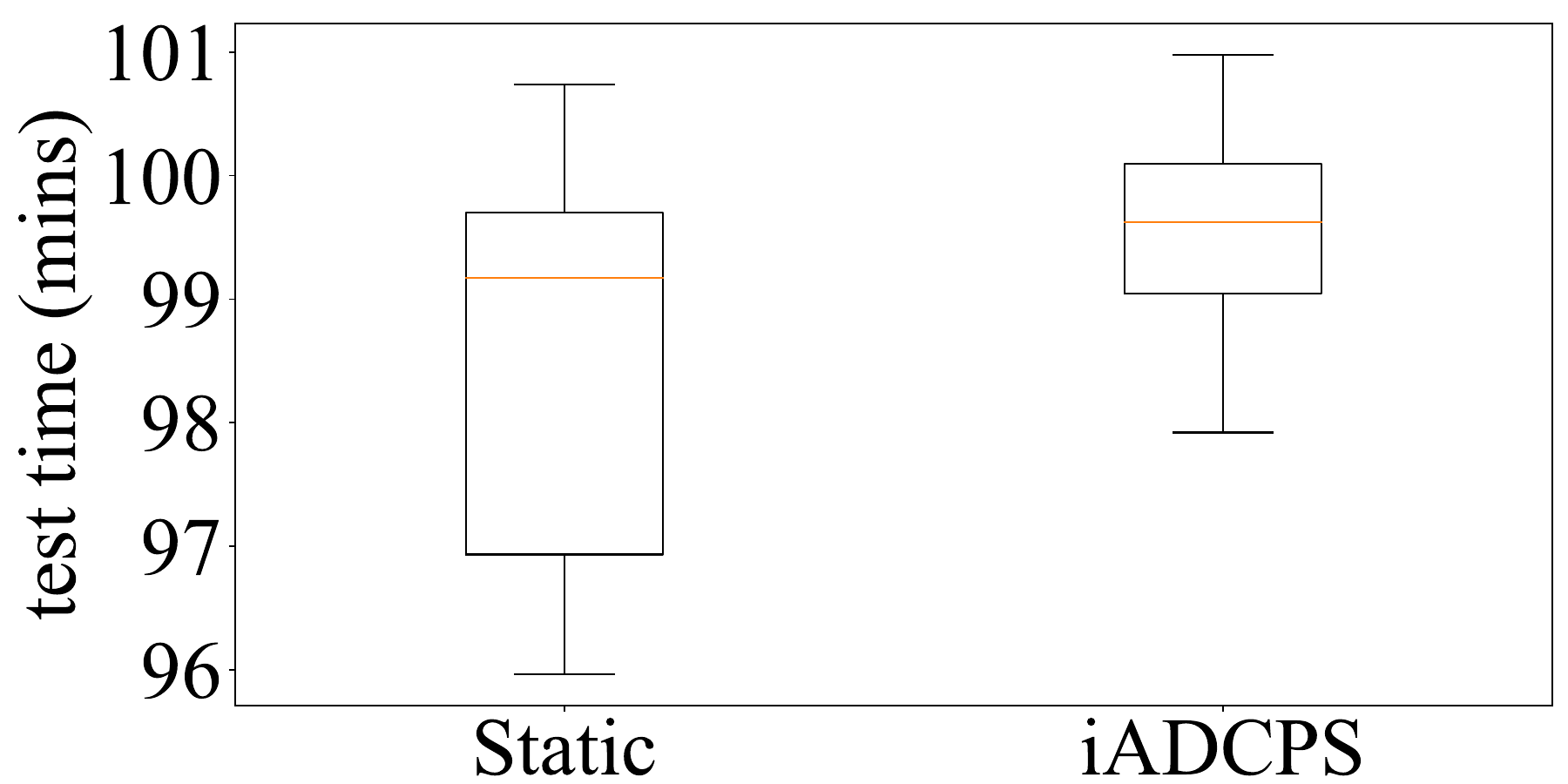}
	}

	\centering
	\scriptsize
	\subfloat[SWaT Training Time]{
		\includegraphics[width=0.48\linewidth]{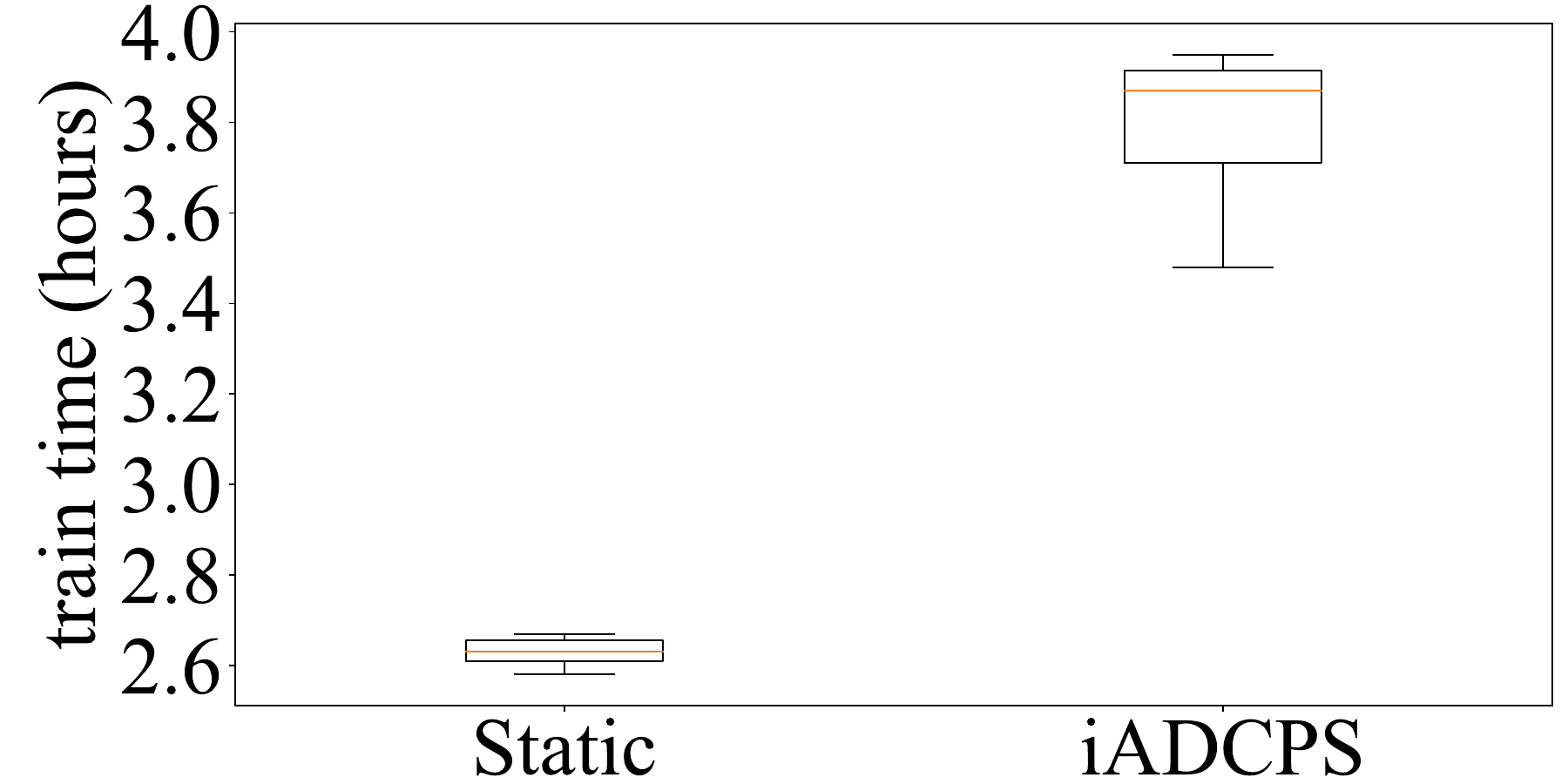}
	}
	\hfill
	\subfloat[SWaT Testing Time ]{
		\includegraphics[width=0.48\linewidth]{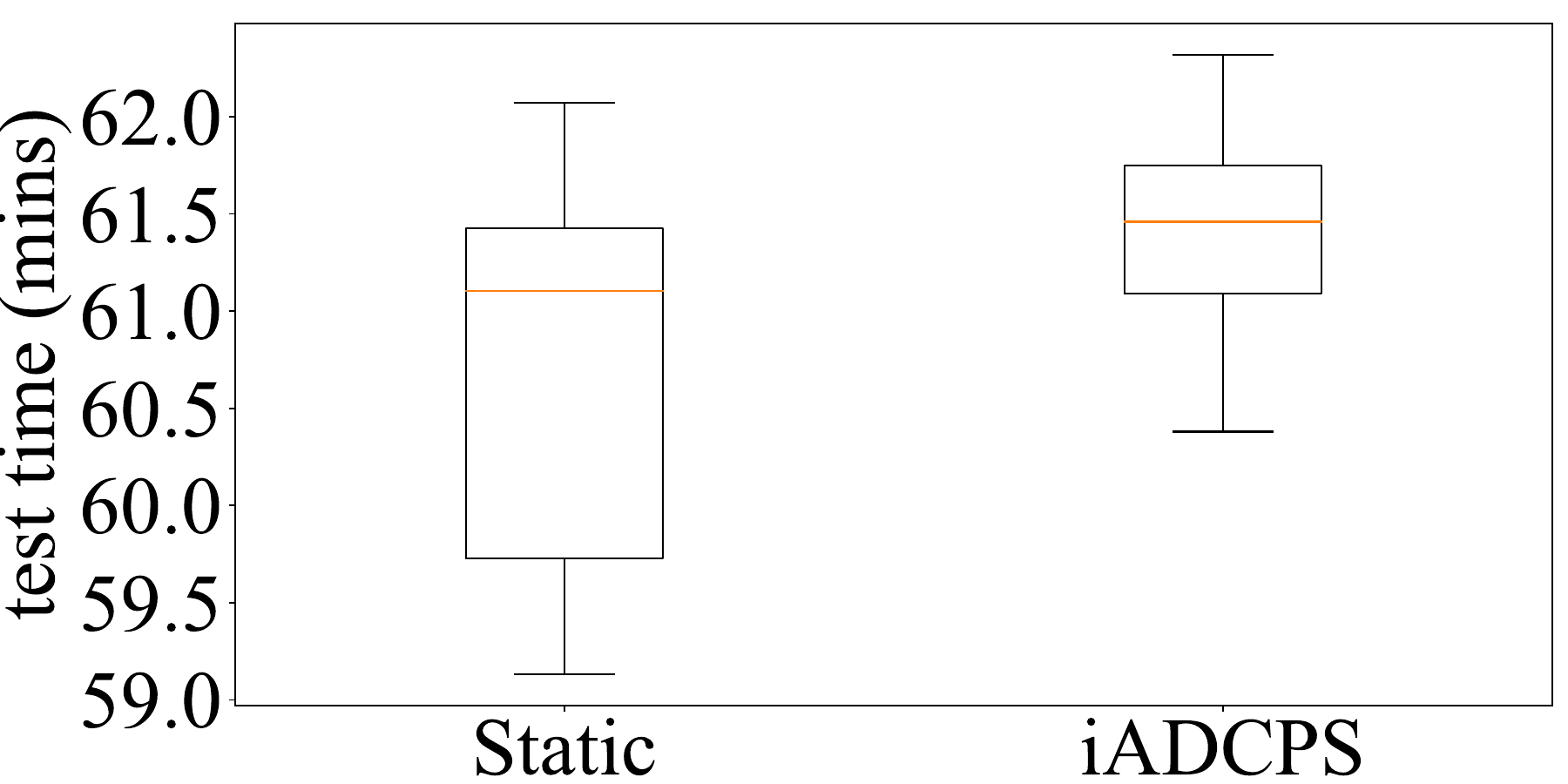}
	}

	\centering
	\scriptsize
	\subfloat[WADI Training Time]{
		\includegraphics[width=0.48\linewidth]{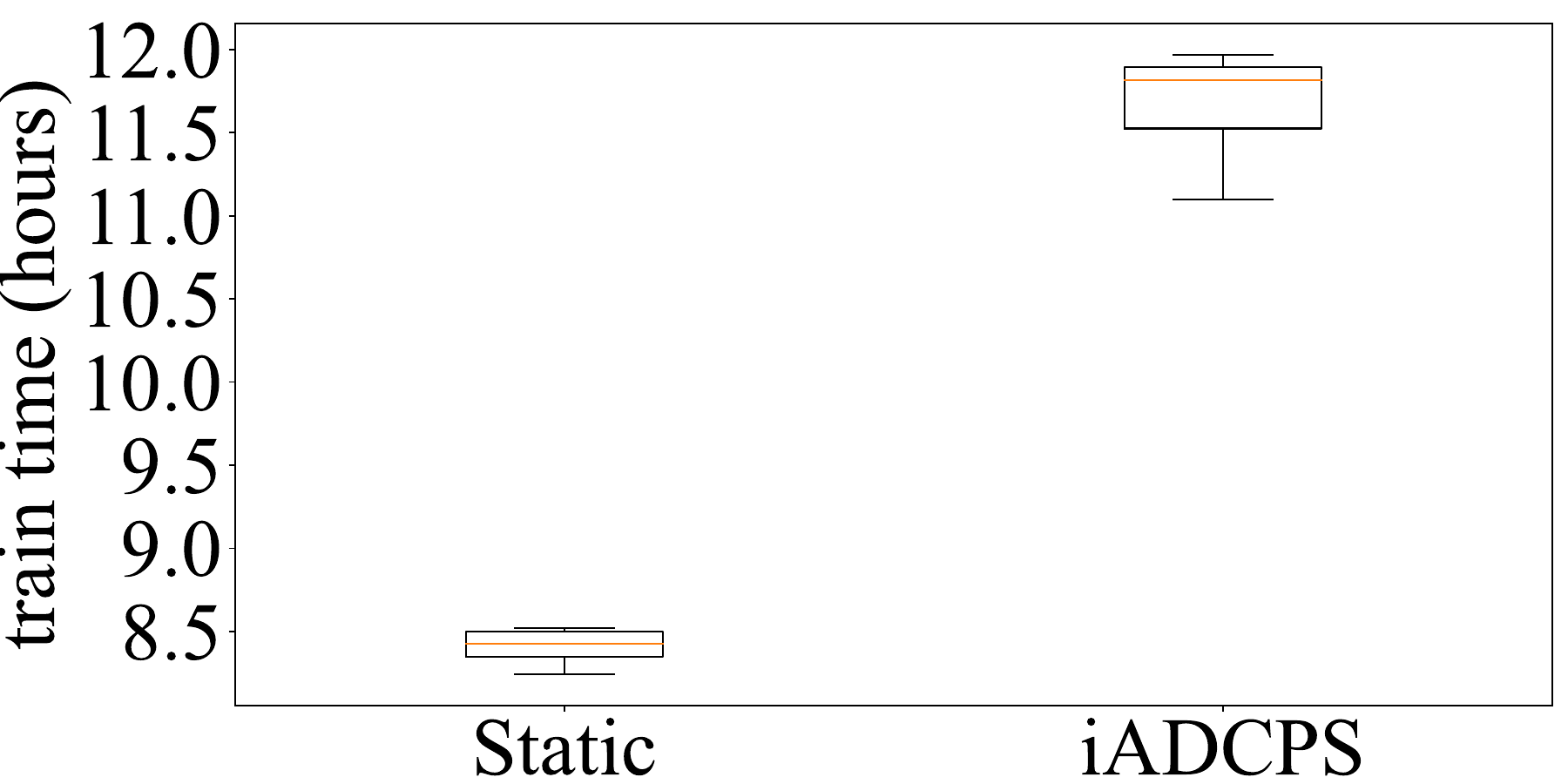}
	}
	\hfill
	\subfloat[WADI Testing Time ]{
		\includegraphics[width=0.48\linewidth]{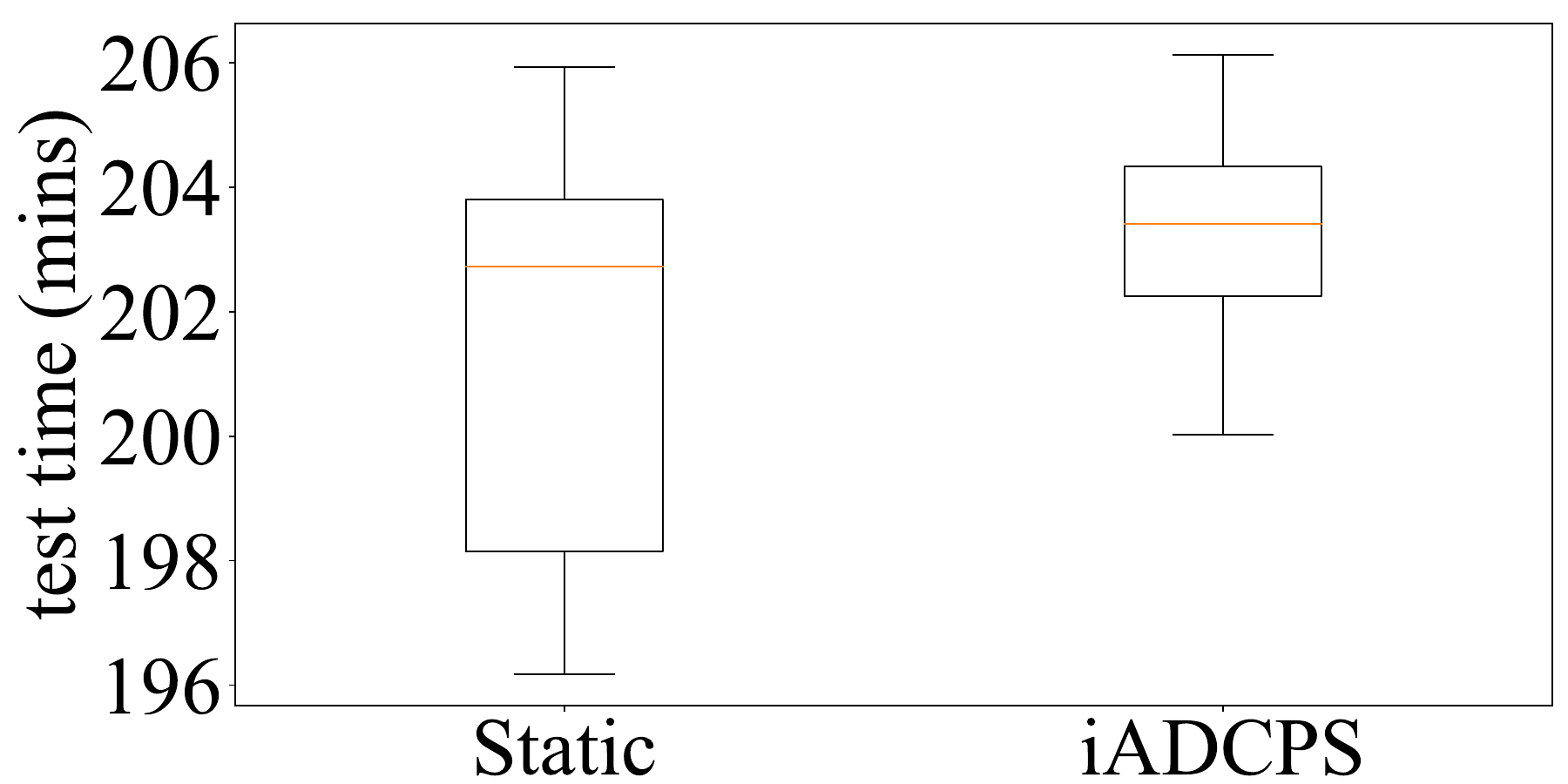}
	}

\caption{Detection time for incremental training and dynamic thresholding.}
\label{figure8}
\end{figure}

We explore the detection efficiency of the proposed method on three Mix datasets, including PUMP, SWaT, and WADI, with specific measures of training time (the sum of initial and incremental training time) and detection time (covering feature extraction time, model detection. Given that the static method does not need to dynamically adjust the threshold according to the anomaly scores, we specifically compare the test time of the static method with that of iADCPS. Here, the static method is implemented using NSIBF with the same end-to-end SSM structure as iADCPS. Figure \ref{figure8} visualizes the comparison results of training time and detection time.

In terms of training time, the main difference between the static method and iADCPS is reflected in the incremental training session. Since the static method does not involve incremental training, its average training time on the PUMP, SWaT, and WADI datasets are 2.5, 2.6, and 8.4 hours, respectively. In comparison, the training times of iADCPS on the same datasets are 3.7, 3.8, and 11.5 hours, respectively. It is worth noting that despite this difference, iADCPS maintains efficient training because incremental training requires far fewer samples than initial training.

In terms of detection time, there is a slight difference between the static and dynamic methods. The detection time of the dynamic method is slightly longer than that of the static method because the threshold needs to be calculated in real-time. However, this difference is not significant for two reasons: firstly, the time spent on feature extraction and model detection in the detection process is much longer than the threshold judgement; secondly, iADCPS limits the number of anomaly scores used for dynamic threshold calculation through the Memory mechanism, which improves the efficiency of threshold calculation.


\subsection{RQ4. Ablation Experiment}
To demonstrate the effect of the main components of iADCPS on the detection performance, we conducted an ablation study under different evolution scenarios. We determine the role of the main components, Dual-Adapter and LDP-DT, by four combinations of the: 1) predefined static models and static thresholds (ST); 2) predefined static models and LDP-DT (LDP-DT); 3) Dual-Adapter and predefined static thresholds (Dual-Adapter+ST); 4) Dual-Adapter and LDP-DT (iADCPS).

\begin{table*}[ht]
\newcommand{\tabincell}[2]{\begin{tabular}{@{}#1@{}}#2\end{tabular}}
\centering
\begin{threeparttable}
\setlength\tabcolsep{10pt}
\caption{Results of ablation study}
\begin{tabular}{ccccccccccc}
\toprule  
\multirow{2}*{Scenarios} &\multirow{2}*{Methods} & \multicolumn{3}{c}{PUMP} & \multicolumn{3}{c}{SWaT}& \multicolumn{3}{c}{WADI}\\
\cline{3-11}~ &~ &PRE & REC & F1 & PRE & REC & F1& PRE & REC & F1\\
\midrule
\multirow{4}*{Remove}&ST&98.4&	79.8&	88.1&	64.2&	100&	78.2&	34.4&	26.1&	29.7 \\
~&LDP-DT&98.2&	97.7&	97.9&	82.1&	90.4&	86.1&	44.5&	54.1&	48.8 \\
~&Dual-Adapter+ST&93.1&	87.9&	90.4&	77.2&	85.7&	81.2&	79.0&	61.7&	69.3 \\
~&Dual-Adapter+LDP-DT (Proposed)&100&	98.1&	99.0&	92.3&	89.5&	90.9&	88.2&	64.7&	74.6 \\
\bottomrule
\multirow{4}*{Upgrade}&ST&95.5&	93.1&	94.3&	91.3&	85.8&	88.5&	34.4&	30.6&	32.4 \\
~&LDP-DT&98.8&	97.6&	98.2&	94.5&	84.6&	89.3&	58.3&	60.3&	59.3\\
~&Dual-Adapter+ST&97.0&	93.8&	95.4&	94.3&	84.2&	89.0&	84.3&	63.5&	72.4 \\
~&Dual-Adapter+LDP-DT (Proposed)&100&	98.1&	99.0&	94.6&	85.3&	89.7&	92.7&	65.2&	76.6 \\
\bottomrule
\multirow{4}*{Mix}&ST&94.6&	82.0&	87.9&	63.6&	100&	77.8&	34.3&	25.9&	29.5 \\
~&LDP-DT&98.1&	96.0&	97.0&	77.8&	89.3&	83.2&	44.1&	52.2&	47.8 \\
~&Dual-Adapter+ST&90.0&	86.7&	88.3&	76.2&	84.8&	80.3&	77.5&	60.9&	68.2 \\
~&Dual-Adapter+LDP-DT (Proposed)&99.8&	99.8&	99.8&	96.1&	72.2&	82.5&	86.1&	61.6&	71.8\\
\bottomrule
\end{tabular}
\label{table2}
\end{threeparttable}
\end{table*}

As shown in Table~\ref{table2}, it can be found that the combination of Dual-Adapter and LDP-DT achieves the optimal detection performance, while the combination of Dual-Adapter with static thresholding even leads to a significant degradation of detection performance. For example, in the Mix scenario, the F1 of Dual-Adapter+ST on the PUMP, WADI, and SWAT datasets are 88.3\%, 80.3\%, and 68.2\%, respectively, which makes it difficult to identify anomalies effectively. It is easy to understand that while the dynamic update of the model generalizes the evolved data, the predefined thresholds are not changed accordingly, leading to an increase in the gap between the thresholds and the updated model, and making the detection performance even lower than that of the static model that has not been updated. Meanwhile, the combination of static models and dynamic thresholds outperforms the combination of static models and static thresholds on most scenarios and datasets, proving the effectiveness of the LDP-DT component, i.e., the dynamic threshold algorithm achieves effective adaptive tuning.

Moreover, the combination of again improves the detection performance, demonstrating that continuous updating of the model based on meta-learning methods enables effective generalization of the system evolution at both data and model levels. Furthermore, the combination of Dual-Adapter and LDP-DT further improves the detection performance, demonstrating that Dual-Adapter can achieve effective generalization of system evolution at both data and model levels.

\section{Discussion}
\label{sec:7}
\subsection{Why does iADCPS work?}
iADCPS outperforms SOTA methods for two primary reasons. Firstly, by introducing an incremental meta-learning mechanism, iADCPS can continuously update the model using a limited number of evolving normal samples, thus narrowing the gap between the distribution of the evolved time series and the distribution of the historical time series, and effectively dealing with the problem of distributional shifts resulting from CPS evolution.  Secondly, iADCPS employs a non-parametric dynamic thresholding technique to adjust thresholds based on abnormal score probability densities dynamically. This adaptive thresholding mechanism enhances anomaly detection accuracy without relying on predefined thresholds.

Our study demonstrated the effectiveness of iADCPS in evolving CPS. Nonetheless, iADCPS still has limitations. Continuous model updates for data distribution changes and computation of non-parametric dynamic thresholds may escalate computational demands and resource utilization. For resource-constrained CPS systems, there may be a trade-off between detection performance and resource consumption. In future work, we will explore model compression methods like knowledge distillation to streamline complex models, reducing computational overhead and resource usage while preserving performance levels.

\subsection{Threats to Validity}

\textbf{Data Quality:} Although the PUMP, SWaT, and WADI datasets have been extensively utilized in prior studies and are publicly accessible, their limited temporal scope may obscure significant system evolution patterns within the original data. Our experimental findings stem from simulated analyses applied to these real datasets. Therefore, drawing conclusions based on these datasets alone may introduce bias and does not fully guarantee their generalizability to real-world environments.

\textbf{Tool Comparison:} In our tool evaluation, we employ the IF, AE, LSTM-Pred, NSIBF, LSTM-NDT, and iADCPS methods using Tensorflow. For LSTM-NDT and NSIBF, we implement them using available public source code \cite{lstm_ndt, nsibf}. Regarding the IF, AE, and LSTM-Pred methods, we replicate them based on their original papers, ensuring consistency with reported outcomes. Additionally, we implement the USAD and FuSAGNet methods using the respective open-source code \cite{usad, fusagnet} provided by these approaches, based on Pytorch. However, potential errors or inconsistencies during conversion might impact the fairness and accuracy of our tool comparisons.

\section{Related Work}
\label{sec:8}
\subsection{Anomaly Detection for Cyber-physical System}
Identifying anomalous behavior of Cyber-physical systems from time series data has been an active field \cite{1dcnn, claim, efsa, fslscnn, aecid, kpi, dtcps, MTGFlow, dldit,sensitivehue}. Existing CPS anomaly detection methods can be divided into two categories: static ADCPS methods \cite{OmniAnomaly, gscmad,sarima, vmddt, mtsdcgan,practicalad, thoc, lattice} and dynamic ADCPS methods \cite{lstm_ndt, adt, acudl, dcdetector}.

Static ADCPS methods ignore the continuously changing characteristics of CPSs, they train the detection model on a fixed data distribution and use preset thresholds to determine the anomalies. These methods mainly include OC-SVM \cite{ocsvm}, CC-SAE \cite{ccsae}, FSL-SCNN \cite{fslscnn}, NSIBF \cite{nsibf}, FuSAGNet\cite{fusagnet}, FuGLAD\cite{fuglad} and CutAddPaste \cite{cutaddpaste} . As the typical unsupervised ADCPS method, OC-SVM utilizes a one-class support vector algorithm to train the model and points out the problem of gradual change of sensor values, which OC-SVM cannot handle efficiently. CC-SAE combines clustering with Siamese autoencoder to achieve weakly supervised anomaly detection using a small number of labeled samples. FSL-SCNN builds a meta-learning model using the Siamese convolutional neural network to differentiate between normal and anomalous samples by measuring the distance between the feature representations of two samples. As a robust ADCPS approach, NSIBF captures the dynamics of the CPS through a state-space model and combines it with a Bayesian filtering algorithm to remove the noise, ensuring that anomalies can be effectively detected even in the presence of a complex system with sensor noise. FuSAGNet combines the graph neural network with the sparse autoencoder to learn sparse latent representations and extracts relationships between features by recurrent feature embedding for inclusion in anomaly detection modeling. As a SOTA method, FuGLAD proposes an anomaly detection architecture based on fused graph structure learning with structured prior knowledge; CutAddPaste also augments time series data by generating pseudo-anomalies.

Dynamic ADCPS methods identify anomalies by continuously updating the detection model to fit the differences in the data distribution in combination with constantly changing thresholds. Such methods include LSTM-NDT \cite{lstm_ndt}, ACUDL \cite{acudl}, and ADT \cite{adt}. LSTM-NDT proposes a combination of unsupervised and nonparametric anomaly thresholding for anomaly detection that does not rely on scarce labels or incorrect parameter assumptions. ACUDL represents the implicit correlation between data by constructing a directed graph structure and uses dynamic graphs to perform adaptive updating of the detection model. ADT models the dynamics in anomaly detection thresholding as a Markov decision process that determines the detection results based on a reinforcement learning mechanism and the state of the environment feedback.

Although static ADCPS methods struggle to accommodate unstable time series data, resulting in subpar detection performance. Dynamic ADCPS methods, while mitigating the problem of distributional shifts due to CPS evolution, focus on only one aspect of the model and the threshold, lacking the capability to effectively handle the challenges related to model generalizability and threshold adaptivity induced by CPS evolution.

\subsection{Meta-learning}
Meta-learning is a process of learning general knowledge from many related tasks. This method can enable the model to quickly generalize to new tasks while reducing the need for training samples. At present, work close to our topic includes two types of meta-learning-based tasks: incremental meta-learning \cite{iodml, metafscil, itaml} and one-class meta-learning \cite{umtra, oodmaml, doubleadapter}.

Incremental meta-learning proposes to combine incremental learning with meta-learning. On the one hand, it achieves rapid generalization of models through meta-learning, and on the other hand, it utilizes the regularization mechanism of incremental learning to prevent catastrophic forgetting of historical knowledge by the model. Typical methods include iTAML \cite{itaml} and DoubleAdapt \cite{doubleadapter}. iTAML is not targeted at a single learning task, but optimizes a set of general parameters across all seen tasks, enabling automatic task recognition and rapid updating for specific tasks. DoubleAdapt does not use the forgetting prevention mechanism of incremental learning but treats each incremental learning task as a meta-learning task and performs rapid generalization from both data and model directions to mitigate the impact of distribution changes.

One-class meta-learning learns a binary classifier using only data from one class, including FS-OCC \cite{fsocc} and DeepTime \cite{deeptime}. OC-MAML Based on the MAML framework, an episode sampling strategy is designed to calculate the loss, which can complete the classification task of class imbalance using only a few normal samples. DeepTime proposed a meta-optimization framework for time series data, which uses connected Fourier feature modules to enhance deep time exponential models to learn high-frequency patterns in time series effectively.

Inspired by the work of Double-Adapter and FS-OCC, our study adapts the data of the evolved CPS system from both data and model directions, using a few normal samples to achieve continuous generalization of the detection model.
 
\section{Conclusion}
\label{sec:9}
In this paper, we propose an incremental meta-learning-based anomaly detection method for evolving CPS systems, namely iADCPS, to construct a continuous training framework that can flexibly adapt to the CPS system evolution. By updating the model using a limited number of normal samples, iADCPS effectively filters historical and evolving data, ensuring sustained high detection accuracy even in novel patterns. Furthermore, iADCPS designs non-parametric dynamic thresholding algorithms that can adaptively adjust the threshold along with the model update, while eliminating the dependence on labels. Extensive quantitative and qualitative experiments conducted on simulated data and three publicly available datasets have demonstrated the enhanced generalization capacity of iADCPS. In future work, we will address the detection efficiency of the model for practical deployment, further enhancing the applicability of iADCPS in real-world scenarios.

\bibliography{paperbib.bib}

\begin{thebibliography}{10}
\providecommand{\url}[1]{#1}
\csname url@samestyle\endcsname
\providecommand{\newblock}{\relax}
\providecommand{\bibinfo}[2]{#2}
\providecommand{\BIBentrySTDinterwordspacing}{\spaceskip=0pt\relax}
\providecommand{\BIBentryALTinterwordstretchfactor}{4}
\providecommand{\BIBentryALTinterwordspacing}{\spaceskip=\fontdimen2\font plus
\BIBentryALTinterwordstretchfactor\fontdimen3\font minus \fontdimen4\font\relax}
\providecommand{\BIBforeignlanguage}[2]{{%
\expandafter\ifx\csname l@#1\endcsname\relax
\typeout{** WARNING: IEEEtran.bst: No hyphenation pattern has been}%
\typeout{** loaded for the language `#1'. Using the pattern for}%
\typeout{** the default language instead.}%
\else
\language=\csname l@#1\endcsname
\fi
#2}}
\providecommand{\BIBdecl}{\relax}
\BIBdecl

\bibitem{survey1}
\BIBentryALTinterwordspacing
H.~Kayan, M.~Nunes, O.~Rana, P.~Burnap, and C.~Perera, ``Cybersecurity of industrial cyber-physical systems: A review,'' \emph{ACM Comput. Surv.}, vol.~54, no. 11s, sep 2022. [Online]. Available: \url{https://doi.org/10.1145/3510410}
\BIBentrySTDinterwordspacing

\bibitem{survey2}
\BIBentryALTinterwordspacing
Y.~Luo, Y.~Xiao, L.~Cheng, G.~Peng, and D.~D. Yao, ``Deep learning-based anomaly detection in cyber-physical systems: Progress and opportunities,'' \emph{ACM Comput. Surv.}, vol.~54, no.~5, may 2021. [Online]. Available: \url{https://doi.org/10.1145/3453155}
\BIBentrySTDinterwordspacing

\bibitem{survey3}
\BIBentryALTinterwordspacing
A.~Bl\'{a}zquez-Garc\'{\i}a, A.~Conde, U.~Mori, and J.~A. Lozano, ``A review on outlier/anomaly detection in time series data,'' \emph{ACM Comput. Surv.}, vol.~54, no.~3, apr 2021. [Online]. Available: \url{https://doi.org/10.1145/3444690}
\BIBentrySTDinterwordspacing

\bibitem{dctgan}
Y.~Li, X.~Peng, J.~Zhang, Z.~Li, and M.~Wen, ``Dct-gan: Dilated convolutional transformer-based gan for time series anomaly detection,'' \emph{IEEE Transactions on Knowledge and Data Engineering}, vol.~35, no.~4, pp. 3632--3644, 2023.

\bibitem{amcnnlstm}
Y.~Liu, S.~Garg, J.~Nie, Y.~Zhang, Z.~Xiong, J.~Kang, and M.~S. Hossain, ``Deep anomaly detection for time-series data in industrial iot: A communication-efficient on-device federated learning approach,'' \emph{IEEE Internet of Things Journal}, vol.~8, no.~8, pp. 6348--6358, 2021.

\bibitem{lstm_vae}
J.~Goh, S.~Adepu, M.~Tan, and Z.~S. Lee, ``Anomaly detection in cyber physical systems using recurrent neural networks,'' in \emph{2017 IEEE 18th International Symposium on High Assurance Systems Engineering (HASE)}, 2017, pp. 140--145.

\bibitem{cgcd}
W.~Tang, J.~Liu, Y.~Zhou, and Z.~Ding, ``Causality-guided counterfactual debiasing for anomaly detection of cyber-physical systems,'' \emph{IEEE Transactions on Industrial Informatics}, vol.~20, no.~3, pp. 4582--4593, 2024.

\bibitem{nsibf}
\BIBentryALTinterwordspacing
C.~Feng and P.~Tian, ``Time series anomaly detection for cyber-physical systems via neural system identification and bayesian filtering,'' in \emph{Proceedings of the 27th ACM SIGKDD Conference on Knowledge Discovery \& Data Mining}, ser. KDD '21.\hskip 1em plus 0.5em minus 0.4em\relax New York, NY, USA: Association for Computing Machinery, 2021, p. 2858–2867. [Online]. Available: \url{https://doi.org/10.1145/3447548.3467137}
\BIBentrySTDinterwordspacing

\bibitem{fusagnet}
\BIBentryALTinterwordspacing
S.~Han and S.~S. Woo, ``Learning sparse latent graph representations for anomaly detection in multivariate time series,'' in \emph{Proceedings of the 28th ACM SIGKDD Conference on Knowledge Discovery and Data Mining}, ser. KDD '22.\hskip 1em plus 0.5em minus 0.4em\relax New York, NY, USA: Association for Computing Machinery, 2022, p. 2977–2986. [Online]. Available: \url{https://doi.org/10.1145/3534678.3539117}
\BIBentrySTDinterwordspacing

\bibitem{lstm_ndt}
\BIBentryALTinterwordspacing
K.~Hundman, V.~Constantinou, C.~Laporte, I.~Colwell, and T.~Soderstrom, ``Detecting spacecraft anomalies using lstms and nonparametric dynamic thresholding,'' in \emph{Proceedings of the 24th ACM SIGKDD International Conference on Knowledge Discovery \& Data Mining}, ser. KDD '18.\hskip 1em plus 0.5em minus 0.4em\relax New York, NY, USA: Association for Computing Machinery, 2018, p. 387–395. [Online]. Available: \url{https://doi.org/10.1145/3219819.3219845}
\BIBentrySTDinterwordspacing

\bibitem{arcus}
\BIBentryALTinterwordspacing
S.~Yoon, Y.~Lee, J.-G. Lee, and B.~S. Lee, ``Adaptive model pooling for online deep anomaly detection from a complex evolving data stream,'' \emph{Proceedings of the 28th ACM SIGKDD Conference on Knowledge Discovery and Data Mining}, 2022. [Online]. Available: \url{https://api.semanticscholar.org/CorpusID:249605470}
\BIBentrySTDinterwordspacing

\bibitem{acudl}
L.~Xi, D.~Miao, M.~Li, R.~Wang, H.~Liu, and X.~Huang, ``Adaptive-correlation-aware unsupervised deep learning for anomaly detection in cyber-physical systems,'' \emph{IEEE Transactions on Dependable and Secure Computing}, vol.~21, no.~4, pp. 2888--2899, 2024.

\bibitem{adt}
\BIBentryALTinterwordspacing
X.~Yang, E.~Howley, and M.~Schukat, ``Adt: Time series anomaly detection for cyber-physical systems via deep reinforcement learning,'' \emph{Computers \& Security}, vol. 141, p. 103825, 2024. [Online]. Available: \url{https://www.sciencedirect.com/science/article/pii/S0167404824001263}
\BIBentrySTDinterwordspacing

\bibitem{ssmkf}
J.~Vilà-Valls, E.~Chaumette, F.~Vincent, and P.~Closas, ``Robust linearly constrained kalman filter for general mismatched linear state-space models,'' \emph{IEEE Transactions on Automatic Control}, vol.~67, no.~12, pp. 6794--6801, 2022.

\bibitem{ccsae}
A.~Castellani, S.~Schmitt, and S.~Squartini, ``Real-world anomaly detection by using digital twin systems and weakly supervised learning,'' \emph{IEEE Transactions on Industrial Informatics}, vol.~17, no.~7, pp. 4733--4742, 2021.

\bibitem{aecid}
F.~Skopik, M.~Wurzenberger, G.~Höld, M.~Landauer, and W.~Kuhn, ``Behavior-based anomaly detection in log data of physical access control systems,'' \emph{IEEE Transactions on Dependable and Secure Computing}, vol.~20, no.~4, pp. 3158--3175, 2023.

\bibitem{kalmanfilter}
D.~Simon, ``Kalman filtering,'' \emph{Embedded systems programming}, vol.~14, no.~6, pp. 72--79, 2001.

\bibitem{cotmix}
E.~Eldele, M.~Ragab, Z.~Chen, M.~Wu, C.~Kwoh, and X.~Li, ``Contrastive domain adaptation for time-series via temporal mixup,'' \emph{IEEE Transactions on Artificial Intelligence}, vol.~5, no.~03, pp. 1185--1194, mar 2024.

\bibitem{maml}
C.~Finn, P.~Abbeel, and S.~Levine, ``Model-agnostic meta-learning for fast adaptation of deep networks,'' in \emph{Proceedings of the 34th International Conference on Machine Learning - Volume 70}, ser. ICML'17.\hskip 1em plus 0.5em minus 0.4em\relax JMLR.org, 2017, p. 1126–1135.

\bibitem{ada}
Y.~Yuan, S.~Srikant~Adhatarao, M.~Lin, Y.~Yuan, Z.~Liu, and X.~Fu, ``Ada: Adaptive deep log anomaly detector,'' in \emph{IEEE INFOCOM 2020 - IEEE Conference on Computer Communications}, 2020, pp. 2449--2458.

\bibitem{if2008}
F.~T. Liu, K.~M. Ting, and Z.-H. Zhou, ``Isolation forest,'' in \emph{2008 eighth ieee international conference on data mining}.\hskip 1em plus 0.5em minus 0.4em\relax IEEE, 2008, pp. 413--422.

\bibitem{usad}
\BIBentryALTinterwordspacing
J.~Audibert, P.~Michiardi, F.~Guyard, S.~Marti, and M.~A. Zuluaga, ``Usad: Unsupervised anomaly detection on multivariate time series,'' in \emph{Proceedings of the 26th ACM SIGKDD International Conference on Knowledge Discovery \& Data Mining}, ser. KDD '20.\hskip 1em plus 0.5em minus 0.4em\relax New York, NY, USA: Association for Computing Machinery, 2020, p. 3395–3404. [Online]. Available: \url{https://doi.org/10.1145/3394486.3403392}
\BIBentrySTDinterwordspacing

\bibitem{fuglad}
S.~He, G.~Li, K.~Xie, and P.~K. Sharma, ``Fusion graph structure learning-based multivariate time series anomaly detection with structured prior knowledge,'' \emph{IEEE Transactions on Information Forensics and Security}, vol.~19, pp. 8760--8772, 2024.

\bibitem{ctad}
H.~Kim, S.~Kim, S.~Min, and B.~Lee, ``Contrastive time-series anomaly detection,'' \emph{IEEE Transactions on Knowledge and Data Engineering}, vol.~36, no.~10, pp. 5053--5065, 2024.

\bibitem{cutaddpaste}
\BIBentryALTinterwordspacing
R.~Wang, X.~Mou, R.~Yang, K.~Gao, P.~Liu, C.~Liu, T.~Wo, and X.~Liu, ``Cutaddpaste: Time series anomaly detection by exploiting abnormal knowledge,'' in \emph{Proceedings of the 30th ACM SIGKDD Conference on Knowledge Discovery and Data Mining}, ser. KDD '24.\hskip 1em plus 0.5em minus 0.4em\relax New York, NY, USA: Association for Computing Machinery, 2024, p. 3176–3187. [Online]. Available: \url{https://doi.org/10.1145/3637528.3671739}
\BIBentrySTDinterwordspacing

\bibitem{1dcnn}
S.~Liu, B.~Zhou, Q.~Ding, B.~Hooi, Z.~Zhang, H.~Shen, and X.~Cheng, ``Time series anomaly detection with adversarial reconstruction networks,'' \emph{IEEE Transactions on Knowledge and Data Engineering}, vol.~35, no.~4, pp. 4293--4306, 2023.

\bibitem{claim}
R.~Wu and E.~J. Keogh, ``Current time series anomaly detection benchmarks are flawed and are creating the illusion of progress (extended abstract),'' in \emph{2022 IEEE 38th International Conference on Data Engineering (ICDE)}, 2022, pp. 1479--1480.

\bibitem{efsa}
L.~Cheng, K.~Tian, D.~D. Yao, L.~Sha, and R.~A. Beyah, ``Checking is believing: Event-aware program anomaly detection in cyber-physical systems,'' \emph{IEEE Transactions on Dependable and Secure Computing}, vol.~18, no.~2, pp. 825--842, 2021.

\bibitem{fslscnn}
X.~Zhou, W.~Liang, S.~Shimizu, J.~Ma, and Q.~Jin, ``Siamese neural network based few-shot learning for anomaly detection in industrial cyber-physical systems,'' \emph{IEEE Transactions on Industrial Informatics}, vol.~17, no.~8, pp. 5790--5798, 2021.

\bibitem{kpi}
H.~Zhu, S.~Rho, S.~Liu, and F.~Jiang, ``Learning spatial graph structure for multivariate kpi anomaly detection in large-scale cyber-physical systems,'' \emph{IEEE Transactions on Instrumentation and Measurement}, vol.~72, pp. 1--16, 2023.

\bibitem{dtcps}
J.~Ma, Y.~Guo, C.~Fang, and Q.~Zhang, ``Digital-twin-based cps anomaly diagnosis and security defense countermeasure recommendation,'' \emph{IEEE Internet of Things Journal}, vol.~11, no.~10, pp. 18\,726--18\,738, 2024.

\bibitem{MTGFlow}
Q.~Zhou, S.~He, H.~Liu, J.~Chen, and W.~Meng, ``Label-free multivariate time series anomaly detection,'' \emph{IEEE Transactions on Knowledge and Data Engineering}, vol.~36, no.~7, pp. 3166--3179, 2024.

\bibitem{dldit}
D.~He, X.~Lv, X.~Xu, S.~Chan, and K.-K.~R. Choo, ``Double-layer detection of internal threat in enterprise systems based on deep learning,'' \emph{IEEE Transactions on Information Forensics and Security}, vol.~19, pp. 4741--4751, 2024.

\bibitem{sensitivehue}
Y.~Feng, W.~Zhang, Y.~Fu, W.~Jiang, J.~Zhu, and W.~Ren, ``Sensitivehue: Multivariate time series anomaly detection by enhancing the sensitivity to normal patterns,'' in \emph{Proceedings of the IEEE/ACM 42nd International Conference on Software Engineering Workshops}, ser. KDD '24.\hskip 1em plus 0.5em minus 0.4em\relax New York, NY, USA: Association for Computing Machinery, 2024, p. 782–793.

\bibitem{OmniAnomaly}
\BIBentryALTinterwordspacing
Y.~Su, Y.~Zhao, C.~Niu, R.~Liu, W.~Sun, and D.~Pei, ``Robust anomaly detection for multivariate time series through stochastic recurrent neural network,'' in \emph{Proceedings of the 25th ACM SIGKDD International Conference on Knowledge Discovery \& Data Mining}, ser. KDD '19.\hskip 1em plus 0.5em minus 0.4em\relax New York, NY, USA: Association for Computing Machinery, 2019, p. 2828–2837. [Online]. Available: \url{https://doi.org/10.1145/3292500.3330672}
\BIBentrySTDinterwordspacing

\bibitem{gscmad}
Z.~Zhang, Z.~Geng, and Y.~Han, ``Graph structure change-based anomaly detection in multivariate time series of industrial processes,'' \emph{IEEE Transactions on Industrial Informatics}, vol.~20, no.~4, pp. 6457--6466, 2024.

\bibitem{sarima}
W.~Hao, T.~Yang, and Q.~Yang, ``Hybrid statistical-machine learning for real-time anomaly detection in industrial cyber–physical systems,'' \emph{IEEE Transactions on Automation Science and Engineering}, vol.~20, no.~1, pp. 32--46, 2023.

\bibitem{vmddt}
V.~K. Singh and M.~Govindarasu, ``A cyber-physical anomaly detection for wide-area protection using machine learning,'' \emph{IEEE Transactions on Smart Grid}, vol.~12, no.~4, pp. 3514--3526, 2021.

\bibitem{mtsdcgan}
\BIBentryALTinterwordspacing
H.~Sun, Y.~Huang, L.~Han, C.~Fu, H.~Liu, and X.~Long, ``Mts-dvgan: Anomaly detection in cyber-physical systems using a dual variational generative adversarial network,'' \emph{Computers \& Security}, vol. 139, p. 103570, 2024. [Online]. Available: \url{https://www.sciencedirect.com/science/article/pii/S0167404823004807}
\BIBentrySTDinterwordspacing

\bibitem{practicalad}
\BIBentryALTinterwordspacing
A.~Abdulaal, Z.~Liu, and T.~Lancewicki, ``Practical approach to asynchronous multivariate time series anomaly detection and localization,'' in \emph{Proceedings of the 27th ACM SIGKDD Conference on Knowledge Discovery \& Data Mining}, ser. KDD '21.\hskip 1em plus 0.5em minus 0.4em\relax New York, NY, USA: Association for Computing Machinery, 2021, p. 2485–2494. [Online]. Available: \url{https://doi.org/10.1145/3447548.3467174}
\BIBentrySTDinterwordspacing

\bibitem{thoc}
L.~Shen, Z.~Li, and J.~T. Kwok, ``Timeseries anomaly detection using temporal hierarchical one-class network,'' in \emph{Proceedings of the 34th International Conference on Neural Information Processing Systems}, ser. NIPS '20.\hskip 1em plus 0.5em minus 0.4em\relax Red Hook, NY, USA: Curran Associates Inc., 2020.

\bibitem{lattice}
\BIBentryALTinterwordspacing
Q.~Xu, S.~Ali, and T.~Yue, ``Digital twin-based anomaly detection with curriculum learning in cyber-physical systems,'' \emph{ACM Trans. Softw. Eng. Methodol.}, vol.~32, no.~5, jul 2023. [Online]. Available: \url{https://doi.org/10.1145/3582571}
\BIBentrySTDinterwordspacing

\bibitem{dcdetector}
\BIBentryALTinterwordspacing
Y.~Yang, C.~Zhang, T.~Zhou, Q.~Wen, and L.~Sun, ``Dcdetector: Dual attention contrastive representation learning for time series anomaly detection,'' in \emph{Proceedings of the 29th ACM SIGKDD Conference on Knowledge Discovery and Data Mining}, ser. KDD '23.\hskip 1em plus 0.5em minus 0.4em\relax New York, NY, USA: Association for Computing Machinery, 2023, p. 3033–3045. [Online]. Available: \url{https://doi.org/10.1145/3580305.3599295}
\BIBentrySTDinterwordspacing

\bibitem{ocsvm}
J.~Inoue, Y.~Yamagata, Y.~Chen, C.~M. Poskitt, and J.~Sun, ``Anomaly detection for a water treatment system using unsupervised machine learning,'' in \emph{2017 IEEE international conference on data mining workshops (ICDMW)}.\hskip 1em plus 0.5em minus 0.4em\relax IEEE, 2017, pp. 1058--1065.

\bibitem{iodml}
K.~J. Joseph, J.~Rajasegaran, S.~Khan, F.~S. Khan, and V.~N. Balasubramanian, ``Incremental object detection via meta-learning,'' \emph{IEEE Transactions on Pattern Analysis and Machine Intelligence}, vol.~44, no.~12, pp. 9209--9216, 2022.

\bibitem{metafscil}
Z.~Chi, L.~Gu, H.~Liu, Y.~Wang, Y.~Yu, and J.~Tang, ``Metafscil: A meta-learning approach for few-shot class incremental learning,'' in \emph{Proceedings of the IEEE/CVF conference on computer vision and pattern recognition}, 2022, pp. 14\,166--14\,175.

\bibitem{itaml}
J.~Rajasegaran, S.~Khan, M.~Hayat, F.~S. Khan, and M.~Shah, ``itaml: An incremental task-agnostic meta-learning approach,'' in \emph{2020 IEEE/CVF Conference on Computer Vision and Pattern Recognition (CVPR)}, 2020, pp. 13\,585--13\,594.

\bibitem{umtra}
\BIBentryALTinterwordspacing
S.~Khodadadeh, L.~Boloni, and M.~Shah, ``Unsupervised meta-learning for few-shot image classification,'' in \emph{Advances in Neural Information Processing Systems}, H.~Wallach, H.~Larochelle, A.~Beygelzimer, F.~d\textquotesingle Alch\'{e}-Buc, E.~Fox, and R.~Garnett, Eds., vol.~32.\hskip 1em plus 0.5em minus 0.4em\relax Curran Associates, Inc., 2019. [Online]. Available: \url{https://proceedings.neurips.cc/paper_files/paper/2019/file/fd0a5a5e367a0955d81278062ef37429-Paper.pdf}
\BIBentrySTDinterwordspacing

\bibitem{oodmaml}
\BIBentryALTinterwordspacing
T.~Jeong and H.~Kim, ``Ood-maml: Meta-learning for few-shot out-of-distribution detection and classification,'' in \emph{Advances in Neural Information Processing Systems}, H.~Larochelle, M.~Ranzato, R.~Hadsell, M.~Balcan, and H.~Lin, Eds., vol.~33.\hskip 1em plus 0.5em minus 0.4em\relax Curran Associates, Inc., 2020, pp. 3907--3916. [Online]. Available: \url{https://proceedings.neurips.cc/paper_files/paper/2020/file/28e209b61a52482a0ae1cb9f5959c792-Paper.pdf}
\BIBentrySTDinterwordspacing

\bibitem{doubleadapter}
\BIBentryALTinterwordspacing
L.~Zhao, S.~Kong, and Y.~Shen, ``Doubleadapt: A meta-learning approach to incremental learning for stock trend forecasting,'' in \emph{Proceedings of the 29th ACM SIGKDD Conference on Knowledge Discovery and Data Mining}, ser. KDD '23.\hskip 1em plus 0.5em minus 0.4em\relax New York, NY, USA: Association for Computing Machinery, 2023, p. 3492–3503. [Online]. Available: \url{https://doi.org/10.1145/3580305.3599315}
\BIBentrySTDinterwordspacing

\bibitem{fsocc}
\BIBentryALTinterwordspacing
A.~Frikha, D.~Krompa{\ss}, H.-G. Koepken, and V.~Tresp, ``Few-shot one-class classification via meta-learning,'' 2020. [Online]. Available: \url{https://openreview.net/forum?id=B1ltfgSYwS}
\BIBentrySTDinterwordspacing

\bibitem{deeptime}
\BIBentryALTinterwordspacing
G.~Woo, C.~Liu, D.~Sahoo, A.~Kumar, and S.~Hoi, ``Learning deep time-index models for time series forecasting,'' in \emph{Proceedings of the 40th International Conference on Machine Learning}, ser. Proceedings of Machine Learning Research, A.~Krause, E.~Brunskill, K.~Cho, B.~Engelhardt, S.~Sabato, and J.~Scarlett, Eds., vol. 202.\hskip 1em plus 0.5em minus 0.4em\relax PMLR, 23--29 Jul 2023, pp. 37\,217--37\,237. [Online]. Available: \url{https://proceedings.mlr.press/v202/woo23b.html}
\BIBentrySTDinterwordspacing

\end{thebibliography}
\bibliographystyle{IEEEtran}

\vfill
\end{document}